\documentclass[11pt]{article}
\usepackage{abstract}
\usepackage[utf8]{inputenc}
\usepackage{amsxtra}
\usepackage{amsthm}
\usepackage{amssymb}
\usepackage{amsfonts}
\usepackage{graphicx}
\usepackage{rotating}
\usepackage{color}
\usepackage{relsize}
\usepackage{epsfig}
\usepackage{subcaption}
\usepackage{verbatim}
\usepackage[printonlyused]{acronym}
\usepackage{setspace}
\usepackage{epstopdf}
\usepackage{authblk}
\usepackage{sectsty}
\usepackage[colorinlistoftodos]{todonotes}
\usepackage[colorlinks=true,allcolors=black]{hyperref}
\usepackage{textcomp,marvosym}
\usepackage{soul}
\usepackage{algorithm}
\usepackage{algorithmicx}
\usepackage{algpseudocode}
\usepackage{url}
\usepackage{hyperref}
\usetikzlibrary{shapes,arrows,automata,positioning}
\usepackage{mathtools}

\usepackage{booktabs}       
\usepackage{nicefrac}       
\usepackage{microtype}      
\usepackage{lipsum}		
\usepackage{graphicx}
\usepackage{caption}
\usepackage{subcaption}
\usepackage{dcolumn}
\usepackage{bm}
\usepackage{xcolor}
\usepackage{amsmath}
\usepackage{comment} 
\usepackage{amssymb}
\usepackage[T1]{fontenc}
\usepackage{algpseudocode}
\usepackage{algorithm}
\usepackage{authblk}
\usepackage[section]{placeins}
\usepackage{float}
\usepackage{cleveref}

\floatstyle{boxed} 
\restylefloat{figure}

\sectionfont{\fontsize{12}{15}\selectfont}
\subsectionfont{\fontsize{10}{12}\selectfont}

\usepackage{geometry}
\title{Data-Driven Mori-Zwanzig: Approaching a Reduced Order Model for Hypersonic Boundary Layer Transition}

\author{Michael Woodward$^{1,2}$\footnote{Email: mwoodward@math.arizona.edu}, Yifeng Tian$^{2}$, Arvind Mohan$^{2}$, Yen Ting Lin$^{2}$, Christoph Hader$^{3}$,\\ Hermann Fasel$^{3}$, Misha Chertkov$^{1}$, Daniel Livescu$^{2}$}

\affil{1. Graduate Interdisciplinary Program in Applied Mathematics, Department of Mathematics, University of Arizona, Tucson, AZ 85721, USA}
\affil{2. Computer, Computational and Statistical Sciences Division, Los Alamos National Laboratory, Los Alamos, NM 87544}
\affil{3. Department of Aerospace and Mechanical Engineering, University of Arizona, Tucson, AZ 85721, USA}

\begin{document}
\maketitle

\begin{abstract}
In this work, we apply, for the first time to spatially inhomogeneous flows, a recently developed data-driven learning algorithm of Mori-Zwanzig (MZ) operators, which is based on a generalized Koopman’s description of dynamical systems. The MZ formalism provides a mathematically exact procedure for constructing non-Markovian reduced-order models of resolved variables from high-dimensional dynamical systems, where the effects due to the unresolved dynamics are captured in the memory kernel and orthogonal dynamics. The algorithm developed in this work applies Mori's linear projection operator and an SVD based compression to the selection of the resolved variables (equivalently, a low rank approximation of the two time covariance matrices). 
We show that this MZ decomposition not only identifies the same spatio-temporal structures found by DMD, but it can also be used to extract spatio-temporal structures of the hysteresis effects present in the memory kernels. We perform an analysis of these structures in the context of a laminar-turbulent boundary-layer transition flow over a flared cone at Mach 6, and show the dynamical relevance of the memory kernels. Additionally, by including these memory terms learned in our data-driven MZ approach, we show improvement in prediction accuracy over DMD at the same level of truncation and at a similar computational cost. Furthermore, an analysis of the spatio-temporal structures of the MZ operators shows identifiable structures associated with the nonlinear generation of the so-called "hot" streaks on the surface of the flared cone, which have previously been observed in experiments and direct numerical simulations.

       
\end{abstract}

\section{Introduction}
\label{sec:intro}

Data-driven reduced-order modeling (ROM) of complex dynamical systems is a rapidly evolving field which has the potential to tackle notoriously challenging problems in engineering and the physical sciences. Many naturally occurring phenomena, such as turbulent flows, can be characterized as high-dimensional nonlinear dynamical systems that exhibit strong coupling across a broad range of scales. In contrast to simulating the dynamics over all the scales, as is done with Direct Numerical Simulation (DNS), reduced-order models seek to describe the dynamics using a low-dimensional space of variables, referred to as "resolved variables" or observables. ROMs can be used to simulate the dynamics at substantially reduced computational costs as well as provide tractable frameworks for analyzing and understanding the underlying physics. 

Many techniques for obtaining reduced order models have been developed over the years for high dimensional dynamical systems, such as those found in fluid dynamics applications \cite{brunton_kutz_2019}. In simulating turbulence, for example, directly coarse-graining the Navier-Stokes equations can be done with large eddy simulation (LES) and Reynolds-Averaged Navier-Stokes (RANS) \cite{sagaut2006large} which reduce the number of scales that need to be resolved but come at the expense of neglecting nonlinear dynamics that may play an important role in the transitional regime. For high-speed flows in particular, this nonlinear transition regime can cover large parts of the geometry and, therefore, a data-driven ROM also taking into account these stages is crucially needed in order to improve the tools for designing future hypersonic vehicles. Driven by this need and the increased availability of high fidelity simulation and experimental data, many promising data-driven model discovery techniques have emerged. One of the most common techniques in the fluid dynamics community involves extracting proper orthogonal decomposition (POD) modes from data, then projecting the full governing equations onto the linear space spanned by POD modes (for example, via Galerkin projection to obtain a ROM of temporal coefficients \cite{lumley_book_2012}). However, the Galerkin projection has several challenges, such as long-term instability of ROMs~\cite{amsallem2014stability,grimberg2020stability} and presence of spurious states in POD modes not corresponding to the true dynamics of the flow~\cite{aubry1993preserving,akhtar2009stability}. Recently, many new advances in ROM methods have emerged based on mixing machine learning with physics informed approaches which are trained on the ground truth data,  e.g. originating from high-fidelity DNS data \cite{mohan_livescu, mohanNN, woodward2021physics, tian_lles_2022, mohan2021learning}. 

Dynamic Mode Decomposition (DMD) is another popular method developed in the fluid dynamics community \cite{schmid_2010, dmd_book} which is {\it equation-free} method, i.e. finding the temporal coefficients does not require projecting the modes onto the governing equations. The DMD method 
provides an accurate decomposition of complex flows into spatio-temporal coherent structures that can be used for short-time future-state prediction and control \cite{schmid_2010, dmd_book}. Although both POD and DMD are developed as reduced order models, they are most often used as diagnostic tools for the  analysis of large scale coherent structures \cite{taira17_modal_analysis}. 

In this manuscript, we utilize the Mori-Zwanzig approach, introduced in \cite{lin2021datadriven_full, 2021Yifeng-PRF}, that generalizes the approximate Koopmanian learning and shows better performance than DMD and extended DMD (EDMD).
The Mori-Zwanzig (MZ) formalism, developed in statistical mechanics nearly half a century ago to construct reduced-order models for high-dimensional dynamical systems \cite{mori1965transport, zwanzig1973nonlinear}, has recently been theoretically connected to the approximate Koopman learning methods \cite{lin2021datadriven_full}, when using Mori's linear projector. The MZ formalism provides a mathematically exact procedure for constructing non-Markovian reduced-order models of resolved variables from high-dimensional dynamical systems, where the effects due to the unresolved dynamics are captured in the memory kernel and orthogonal dynamics \cite{lin2021datadriven_full}. The Mori-Zwanzig formalism constructs equations describing the evolution of a set of measurable variables, referred to as observables or resolved variables, similar to the Koopmanian description. This MZ formalism provides a mathematically exact procedure for developing reduced-order models for high-dimensional systems, with the result generally depending on its past history. The resulting formulation, referred to as the Generalized Langevin equation (GLE), consists of a Markovian term, a memory term, and a noise term. The Mori–Zwanzig memory term quantifies the interactions between the resolved and under-resolved dynamics, and is related to the noise term through the fluctuation-disspation theorem.  The memory effect depends on the choice of observables and the projection operator. Up until recently, modeling turbulence with the MZ formalism has been extremely challenging due to the unknown structure of the memory kernel, which is affected by the unresolved orthogonal dynamics \cite{2021Yifeng-PRF, duraisamy17_prf}.  However with the recent progress made in \cite{lin2021datadriven_full,lin22_nn_mz}, there are now data-driven methods to learn the Markovian and memory operators in MZ \cite{lin2021datadriven_full, lin22_nn_mz}, with promising results already seen in stationary homogeneous isotropic turbulence \cite{tian_2021}. In this work, we demonstrate the improvement of these new methods over DMD when applied to the complex flow physics present in laminar-turbulent boundary-layer transition in hypersonic boundary layers.

Understanding, predicting and controlling laminar-turbulent boundary-layer transition in hypersonic boundary layers is crucial for the design and safe operation of next generation high-speed vehicles. Transition to turbulence leads to significant increases in skin-friction (drag) and heat transfer and can result in the development of so-called "hot" streaks that can locally far exceed the turbulent heat transfer values (see for example \cite{hader_fasel_2019, meersman_2021}). Therefore, reliable estimates of where transition occurs are vital for predicting aero-thermal loads, surface temperatures and drag during the design stages of a high-speed vehicle. In addition, the development of flow control strategies to either delay or accelerate transition to turbulence of high-speed boundary-layers requires reduced order modeling of the dominant mechanisms leading to transition. Successful flow control strategies could substantially reduce skin friction drag and the weight of the required thermal protection systems.

In this manuscript, we approach these challenging problems by extending the work by \cite{lin2021datadriven_full, lin22_nn_mz, tian_2021}, and developing a data-driven Mori-Zwanzig based ROM for a hypersonic laminar-turbulent boundary-layer transition flow on a flared cone at Mach 6 and zero angle of attack \cite{hader_2018, hader_fasel_2019}. We use Mori's linear projection and SVD based compression for selecting observables (equivalently a low rank approximation of the two time covariance matrices as is done in DMD \cite{dmd_book}, and can be interpreted as projecting onto the POD modes). This data-driven MZ algorithm provides higher-order and memory-dependent corrections to the existing data-driven learning of the approximate Koopman operators using DMD for a similar computational cost.  With this framework, we show that the modes and spectrum obtained from DMD are identical to the modes and spectrum of the Markovian operator of MZ. Furthermore, by including more memory terms in the MZ framework, not only can this data-driven MZ formulation outperform DMD in future state prediction for the hypersonic boundary-layer flow, these memory terms contain nontrivial large scale spatio-temporal structures of the hysteresis effects relevant for the understanding of the transition mechanisms (as higher order corrections for DMD truncation). Thus, we demonstrate that the data-driven MZ approach can serve as a way to understand the Markovian contributions within the linear framework, and offer insights into the hysteresis effects through the non-Markovian structures of the flow, which are dynamically relevant within the transition region.




\section{Koopman Operator and Dynamic Mode Decomposition}
The main data-driven algorithm introduced in this work (see \autoref{sec:algs}) involves similar ideas and concepts used in DMD and the Koopman description of dynamical systems. We give a brief overview of the Koopman operator and the DMD method in order to set up some notations and concepts used later.  The DMD procedure computes the eigendecomposition of the best fit linear (and Markovian) operator $\bm A$ that advances a set of observables $\bm g$ forward in time by the locally linear dynamical system $\bm \dot{\bm g}(t) = \bm A \bm g(t)$ \cite{dmd_book} and is used to approximate the modes of the Koopman operator $\mathcal{K}$ (linear operator acting on infinite dimensional Hilbert space of observables). Similar to other modal decomposition techniques, DMD is most often applied as a diagnostic tool providing physical insight into complex fluid dynamical systems. Since the Koopman operator is linear, it can be characterized by its eigenvalues and eigenfunctions. One advantage of DMD over a Galerkin projection onto POD modes is that DMD is purely data-driven and does not require projecting the governing equations onto the linear subspace spanned by the modes, and instead gives a reduced description of the best fit linear operator that describes the dynamics. 

Consider a discrete time nonlinear autonomous dynamical system, $ \bm x_{n+1} = \mathcal{N}(\bm x_n),$ where $ \mathcal{N}$ is the flow map acting on the states $\bm x$. Then the discrete time Koopman operator $ \mathcal{K} $ is an infinite dimensional linear operator acting on observables $\bm g $, which are functions of the state space variables: $\mathcal{K} \bm g = \bm g \circ \bm x $. Thus, the discrete time Koopman operator defines a new discrete time linear dynamical system, albeit infinite dimensional, that governs the evolution of the observables. When approximating the Koopman operator by a finite dimensional matrix, this essentially provides a linear approximation of a nonlinear system without directly linearizing around a particular fixed point. Based on the Koopman representation of dynamical systems, approximate learning methods, such as dynamic mode decomposition (DMD) \cite{schmid_2010} and extended dynamic mode decomposition (EDMD) have been developed for data-driven modeling of these systems \cite{williams_2015_edmd}. Although the dynamics of the observables are always linearly dependent on other observables, to derive a closed-form solution in the Koopman formulation requires the appropriate identification of a set of observables so that the dynamics are invariant in a subspace which is linearly spanned by the set of observables. This last statement becomes important in the next section, where the MZ approach closes the system by leveraging projection operators, and was shown to be a generalization of the Koopman formulation \cite{lin2021datadriven_full}. 

Based on the approximate Koopmanian learning framework, one aims to construct the evolution of a set of linearly independent observables. 
Since the Koopman operator is linear, it can be characterized by its eigenvalues and eigenfunctions. Given the eigenfunctions, the observables can be expressed as a linear combination of a countably infinite sum of the Koopman eigenfunctions (assuming a discrete spectrum). The DMD method approximates the modes of the Koopman operator.

 
 \subsection{Basic DMD algorithm}
 
The standard DMD procedure produces a low rank eigendecomposition of $\bm A$ that optimally fits the measured trajectory $\bm x_k$ in a least-squares sense, i.e. by minimizing $ ||\bm x_{k+1} - \bm A \bm x_k||_2$. Given the snapshot matrices $\bm X_1 = [\bm x_1, ..., \bm x_{m-1}]$ and $\bm X_2 = [\bm x_2, ..., \bm x_m]$, $ \bm X_2 \approx \bm A \bm X_1$ and the least squares solution is $ \bm A = \bm X_2 \bm X_1^{\dagger}$ where $\dagger$ denotes Moore-Penrose pseudoinverse. However, instead of solving for $\bm A$ directly, as this would require massive amounts of memory for high-dimensional systems, first a low rank approximation $\bm \tilde{\bm A}$ of $\bm A$ is obtained by projecting onto a low rank subspace defined by POD modes.  The following method was shown by \cite{H_Tu_2014} to extract the eigenvectors and eigenvalues of $\bm A$, albeit through the low rank approximation $\tilde{\bm A}$. 

\begin{itemize}
    \item [] Step 1. Truncated SVD $\bm X_1 \approx \bm U_r \bm \Sigma_r \bm V_r^*$ provides low rank truncation ($r$ is level of truncation)
    \item [] Step 2. Project $\bm A$ onto POD modes $\bm \tilde{\bm A} = \bm U_r^* \bm A \bm U_r = \bm U_r^* \bm X_2 \bm V_r \bm \Sigma_r^{-1}$ (we use a similar idea in MZ algorithm \ref{alg:low_rank})
    \item [] Step 3. Compute eigendecomposition $ \tilde{\bm A}\bm W = \bm W \bm \Sigma_r$
    \item [] Step 4. Eigenvectors are given by $\bm \Psi = \bm X_2 \bm V_r \bm \Sigma_r^{-1}\bm W$.
\end{itemize}
 
The extended DMD \cite{williams_2015_edmd}, which is another approximate Koopman learning method, uses a general set of observables $\bm G_1 = [\bm g_1, ..., \bm g_{m-1}]$, and $\bm G_2 = [\bm g_2, ..., \bm g_{m}]$ in place of $\bm X_1$, and $\bm X_2$ and generally includes nonlinear polynomial terms in the state variables $\bm x_k$. The goal here is to develop a set of observables that is rich enough so that the dynamics are invariant in a subspace which is linearly spanned by the set of observables, in order to find an accurate finite rank approximation of the Koopman operator. The two key takeaways here are the Koopmanian description of dynamical systems, and the use of SVD for low rank approximations as both will be utilized in the next section.

\section{Data-Driven Mori-Zwanzig Formulation}
\label{sec:mz_formulation}

In this section we give a brief overview of the mathematical formulation and data-driven learning procedure for an MZ based ROM (see \cite{lin2021datadriven_full} for more detailed discussion and derivation). Y.T. Lin et al. \cite{lin2021datadriven_full} proposed a data-driven learning framework for extracting MZ memory kernel and orthogonal dynamics from high-dimensional data under the generalized Koopman formulation. By combining the Koopman description with the MZ formalism, one can perform a dimensional reduction of the infinite dimensional Koopmanian linear formulation to a finite, low-dimensional dynamical system with memory kernels and orthogonal dynamics. Since the observables evolve in a linear space, the learning problem is convex, which simplifies learning the MZ operators. The final result is a closed dynamical system describing the evolution of observables. However, one should note that the memory and orthogonal dynamics operators contain contributions from the full space of observables. Nevertheless, the algorithms provided in \cite{lin2021datadriven_full} offer a computationally feasible avenue for extracting these operators. 

In contrast to decomposing the observables as an expansion about the Koopman eigenfunctions, the Mori–Zwanzig formalism utilizes the inner product in the Hilbert space (of observables) to decompose the space into the subspace linearly spanned by the set of observables, $\mathcal{H}_{\bm g} := Span\{\mathcal{M}\}$ and an orthogonal subspace $\mathcal{H}_{\bm \bar{\bm g}} := \{ \bar{ g} \in \mathcal{F} : \left< \bar{g}, g_i \right> = 0, g_i \in \mathcal{M} \}$ with a projection operator, where $\mathcal{M} := \{g_i\}_{i=1}^r$.

\subsection{Mori-Zwanzig Formalism}
For the continuous case, consider the autonomous dynamical system evolving the physical variables or states $\bm x$ of the form 
\begin{equation}\label{eq:dynamics}
    \cfrac{d \bm x(t)}{dt} = \bm \Phi (\bm x(t)), \quad \bm x(0) = \bm x_0,
\end{equation}
where $\bm x \in \mathbb{R}^N$, and $\bm \Phi : \mathbb{R}^N \rightarrow \mathbb{R}^N$. Now define the set of observable functions $\bm g : \mathbb{R}^N \rightarrow \mathbb{R}^r$ of the state of the system $\bm x$. These observables are potentially nonlinear and in general $r \ll N$.

The MZ formulation results in the following generalized Langevin equation (GLE) describing the exact evolution of resolved components given an initial condition $\bm x_0$
\begin{equation}\label{eq:gle}
    \frac{\partial}{\partial t} \bm g(\bm x_0, t) =  \bm M (\bm g(\bm x_0, t))  - \int_0^t \bm K(\bm g(\bm x_0, t-s), s) ds + \bm F(\bm x_0, t),
\end{equation}
where $\bm M$, $\bm K$, and $\bm F$ are the Markov, memory and orthogonal dynamics operators, respectively. Details on the derivation can be found in \cite{lin2021datadriven_full}. The above equation is general for any projection operator which maps functions of the full configuration to functions of only the resolved variables in the projected space. Using Mori's linear projection \cite{mori1965transport}, whose projection operator is the functional projection that uses the equipped inner product in the $L^2$ Hilbert space, results in a linear Markovian form $\bm M (\bm g(\bm x_0, t)) = \hat{\bm M} \cdot \bm g(\bm x_0, t)$, and a linear memory dependence $\bm K(\bm g(\bm x_0, t-s), s) = \hat{\bm K}(s) \cdot \bm g(\bm x_0, t-s) $. $\hat{\bm M}  \in \mathbb{R}^{r \times r} $ is termed as the Markovian transition matrix because it depends only on the values of the variables at the current instant, and is identical to the best-fit $\bm A$ in DMD. $\hat{\bm K}  \in \mathbb{R}^{r \times r} $ is termed as the memory kernel, which accounts for the "echo" of resolved observables $\bm g(s)$ at an earlier time $s$ as they affect the under-resolved variables which in turn affect the future evolution of the resolved variables. $\hat{\bm K}$ also accounts for how the initial conditions of the orthogonal observables $\bm g_{\mathcal{\bar{M}}}$ propagate forward to affect the resolved variables. $\bm F$ can be thought of as a noise term, but is formally describing the orthogonal dynamics, which can be as difficult to solve as the full dynamical system in the physical space. In this work, we assume that $\bm F$ is a small residual term and negligible during prediction, however, to further bolster this assumption, nonlinear projection operators by regression to minimize $\bm F$ can be explored in future work, similar to \cite{lin22_nn_mz}. 

In the discrete-time Mori–Zwanzig formulation,  \autoref{eq:dynamics} is replaced with $\bm x_{n+1} = \mathcal{N} (\bm x_n)$, which is more suitable for discrete-time data found from simulations or experiments, where time is discretized as $t = n \Delta$. The resulting generalized Langevin equation (GLE) is 
\begin{equation}\label{eq:gle_discrete}
    \bm g_{n+1} =\bm \Omega^{(0)}_{\Delta} \cdot \bm g_{n} +  \sum_{m=1}^n \bm \Omega^{(m)}_{\Delta} \cdot \bm g_{n-m} + \bm W_n,
\end{equation}
where in practice $k<n$ is the selected number of memory terms to be included (a truncation of the full memory in GLE). Here, $\bm g(\cdot, n\Delta t) = \bm g_n$, and $\bm \Omega^{(m)}_{\Delta}$ are the $\Delta$-dependent, $r\times r$ matrices which are related to the Markovian and memory terms above. For small $\Delta \ll 1$, the relations are the following $\bm \Omega_{\Delta}^{(0)} \approx \bm I + \Delta \hat{\bm M}$ and $\bm \Omega_{\Delta}^{(m)} \approx \Delta^2 \hat{\bm K}(m\Delta)$ when $m>0$. The \textit{noise} term $\bm W_n$, describing the orthogonal dynamics, can be extracted from data once Markovian and memory kernels are learned. However in this work, we focus on learning Markovian and memory kernels and assume the orthogonal dynamics are small during prediction and can be projected out to find the best approximation of the dynamics $\bm g(t)$ in the parallel space.

As with the main use of DMD as a diagnostic tool to understand the underlying physical process, we demonstrate in this work that MZ can also be used to extract spatio-temporal coherent structures, of which the Markovian term captures the DMD modes. 

\subsection{MZ Algorithm}
The algorithm used in this work (see \autoref{sec:algs}) extends what is done in the discrete-time case \cite{lin2021datadriven_full} by using an SVD based low rank approximation, similar to what is done in the DMD procedure, of the otherwise unwieldy covariance matrices involved. Alternatively, this can be interpreted through the lens of SVD as a lossy data compression which is essentially a linear auto-encoder \cite{GoodBengCour16} to select the most energetic observables. 

Here, algorithm \ref{alg:obs} and algorithm \ref{alg:low_rank} are two different interpretations of the same procedure. Algorithm \ref{alg:obs} interprets the SVD as a linear auto-encoder, which is an automatic technique for selecting observables from data (as a lossy data compression to avoid the intractable computations of $\bm X_k \cdot \bm X_1^T$) and algorithm \ref{alg:low_rank} interprets this as a low rank approximation of $\bm C_k$ by projecting onto the POD modes (which is what is done in the standard DMD algorithm).

\section{Results}
First, the algorithm was validated for incompressible flow around a 2D cylinder with $Re=100$ (see \Cref{sec:2d_cyl}). The Markovian modes and the spectrum obtained from MZ match the DMD modes and spectrum. Furthermore, the MZ approach also successfully extracts the spatio-temporal coherent structures similar to what is found with DMD, as well as the spatio-temporal structures of the memory effects (\Cref{sec:2d_cyl}). Next, we investigate the data-driven MZ approach as a ROM for predicting and analyzing high-speed laminar–turbulent boundary-layer transition on a flared cone at Mach 6 with zero angle of attack \cite{hader_2018, hader_fasel_2019}. We show that the memory terms play a significant role in the analysis of the laminar-turbulent transition (Fig. \ref{fig:mz_modes_on_cone}), as well as increasing the accuracy of future state prediction over DMD (Fig. \ref{fig:prediction_comp}).

\subsection{Hypersonic Boundary-Layer Transition}

High-speed laminar–turbulent boundary-layer transition is very complex and remains an active research area in fluid dynamics. As discussed in \cref{sec:intro}, the understanding of high-speed boundary-layer transition is necessary in order to develop reliable transition prediction methods that can be used for the design and safe operation of advanced high-speed vehicles. To the authors’ best knowledge, there have not yet been any studies using data-driven methods to accurately extract MZ terms for studying hypersonic boundary-layer transition. Understanding the memory effects (see \Cref{sec:mz_formulation}), however, may provide crucial understanding for developing reliable transition prediction models. In this work we use high fidelity DNS data to investigate if the data-driven MZ formulation can be used to analyze and obtain a detailed understanding of the dominant transition mechanisms, in particular the nonlinear stages. 

\begin{figure}[!htb]
\centering
\begin{subfigure}[]{0.48\textwidth}
\centering
\includegraphics[width=1\textwidth]{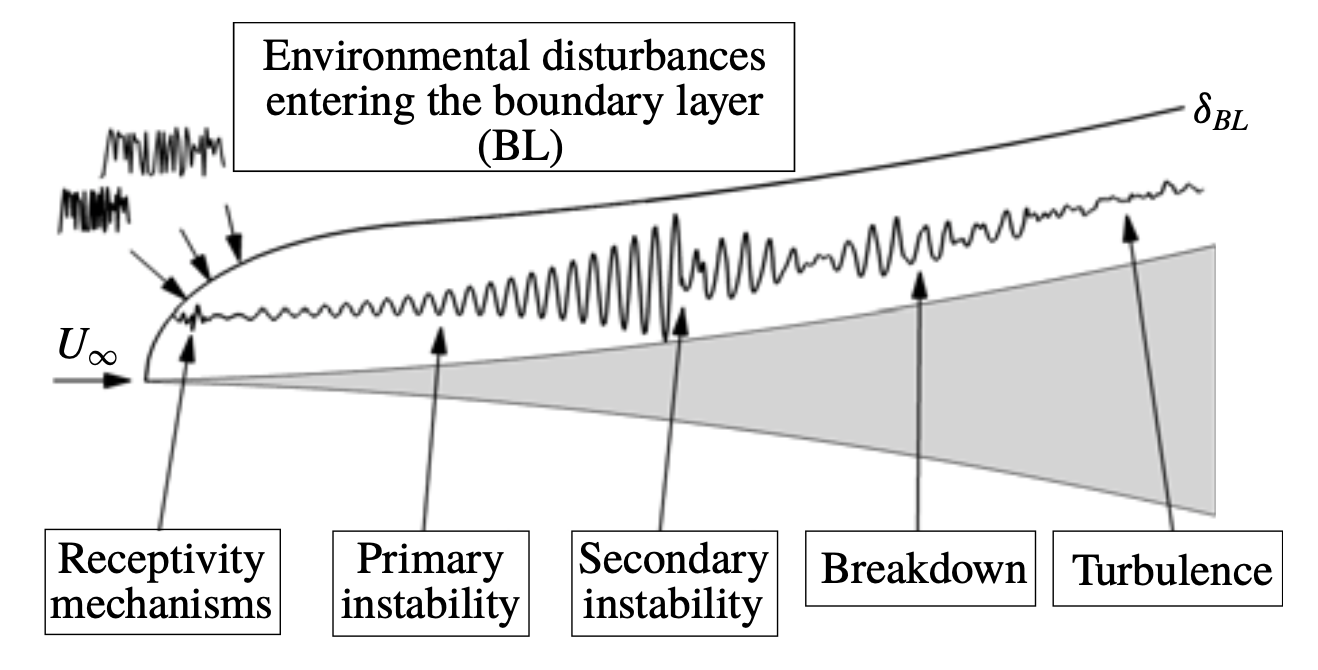}
\caption{}
\label{fig:schematic_path_a}
\end{subfigure}
\begin{subfigure}[]{0.48\textwidth}
\centering
\includegraphics[width=1\textwidth]{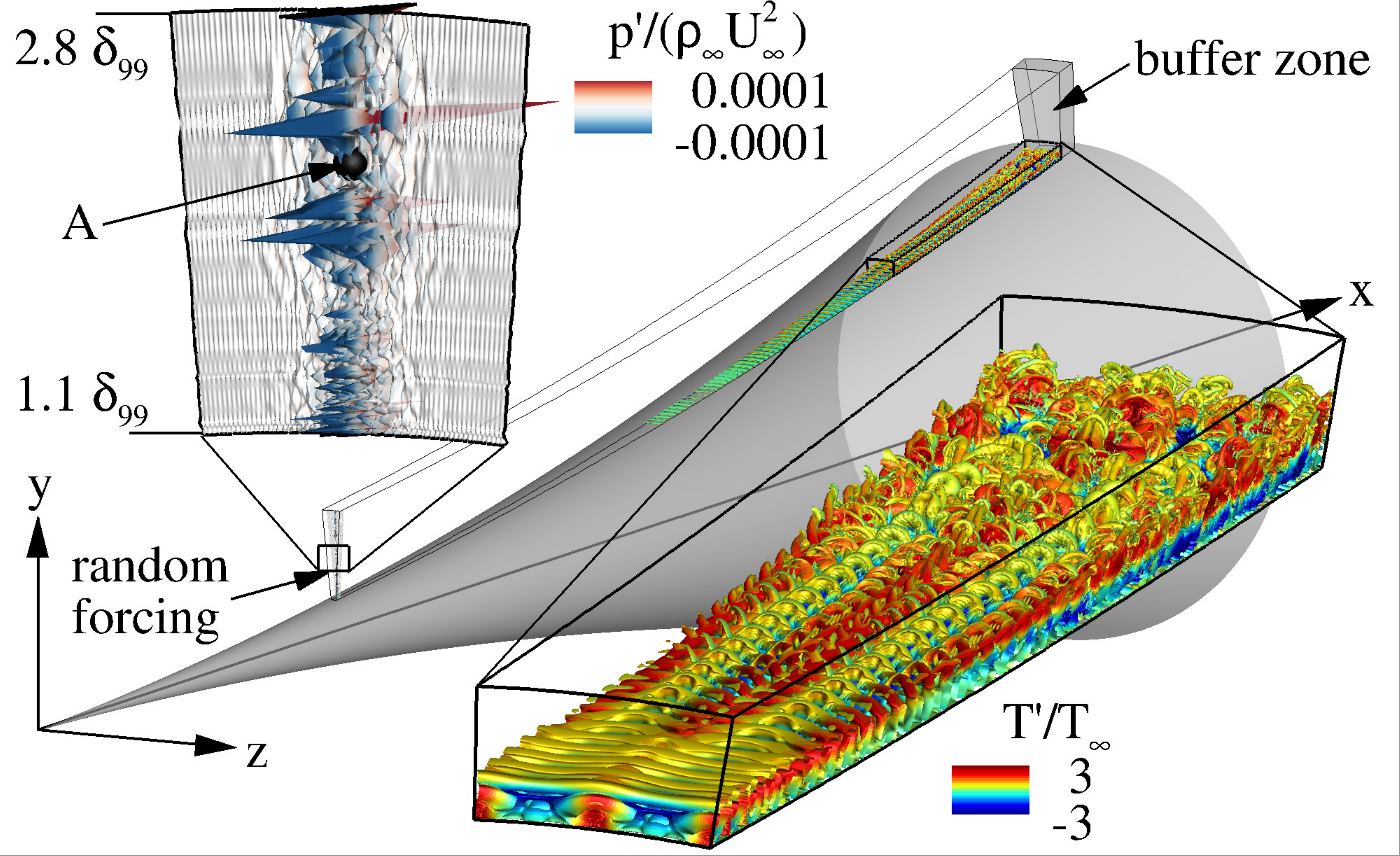}
\caption{}
\label{fig:computational_domain_natural_transition}
\end{subfigure}
\begin{subfigure}[]{0.98\textwidth}
\centering
\includegraphics[width=1\textwidth]{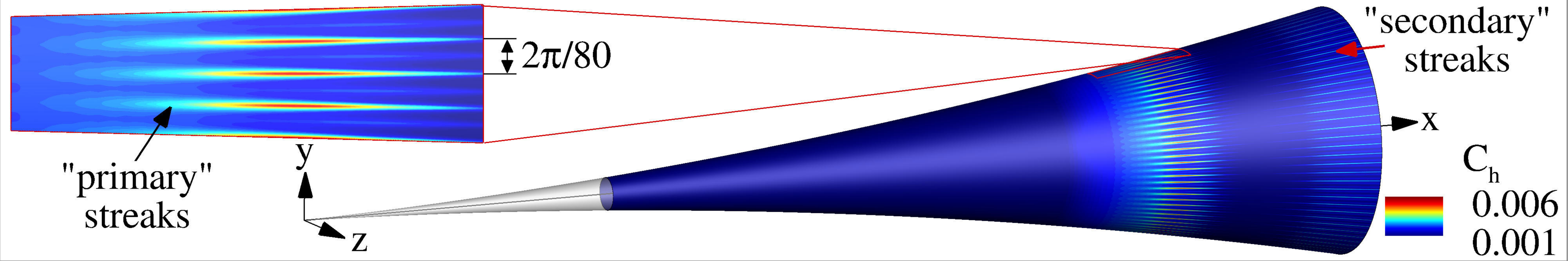}
\caption{}
\label{fig:ch_contours}
\end{subfigure}
\caption{Schematic of the transition stages (\subref{fig:schematic_path_a}), computational setup for the "natural" transition DNS using random forcing (\subref{fig:computational_domain_natural_transition}), and time-averaged Stanton number ($C_h$) contours on the surface of the cone (\subref{fig:ch_contours}).}
\label{fig:geometry}
\end{figure}

We extract and analyze the coherent structures from the Markovian and memory terms from the data-driven MZ formulation applied to data obtained from a DNS of hypersonic boundary layer transition on a flared cone at Mach 6, where transition was initiated by random perturbations at the inflow of the computational domain ("natural" transition, see \cite{hader_2018}). For this DNS, all stages of the so-called path (schematically shown in \Cref{fig:schematic_path_a}) according to the classification by \cite{morkovin_1994} from the linear (primary) instability all the way to turbulence were considered (see \Cref{fig:computational_domain_natural_transition}). 

\begin{figure}[!htb]
\centering
\begin{subfigure}[b]{1.0\textwidth}
\centering
\includegraphics[width=0.95\textwidth]{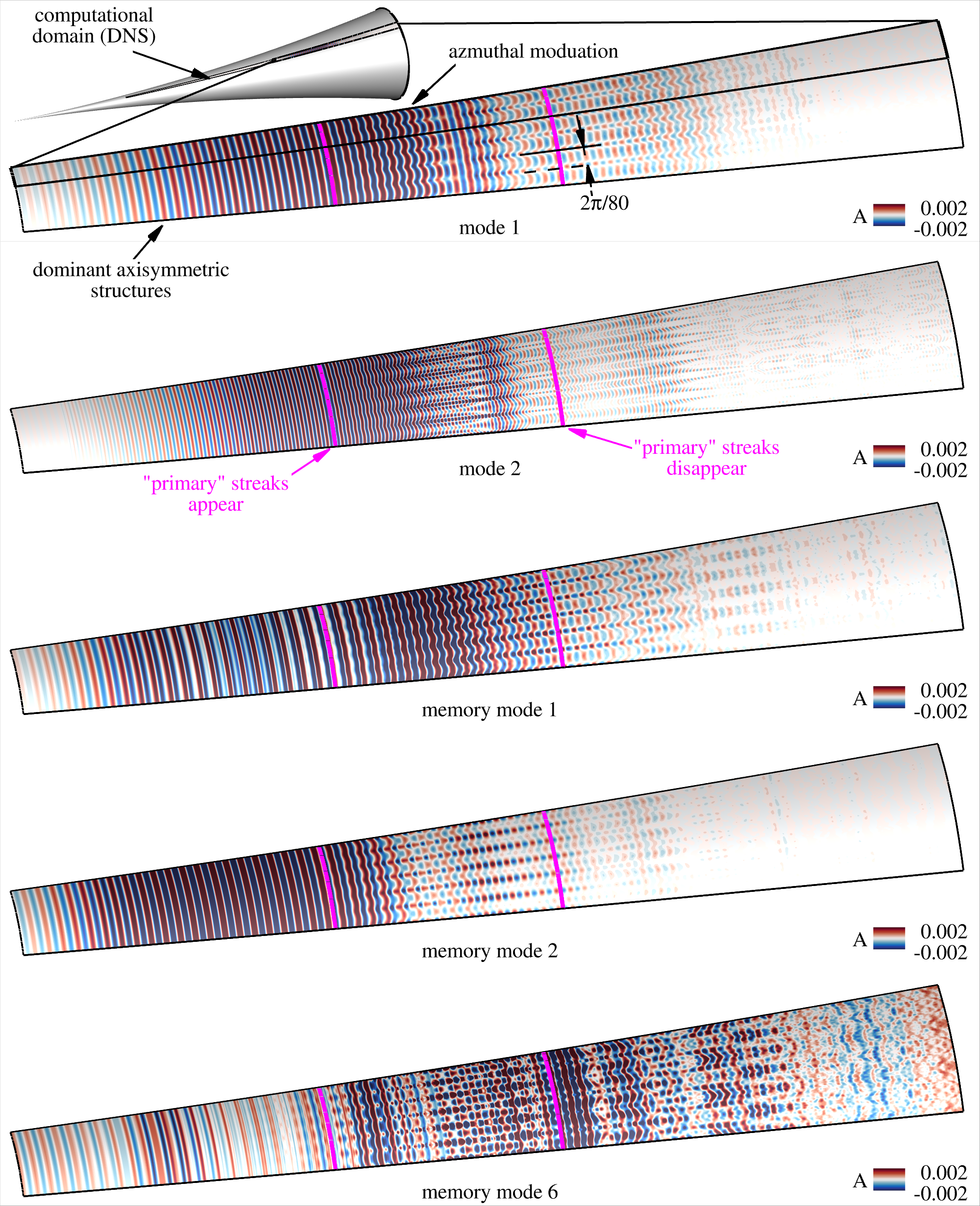}
\end{subfigure}
\caption{MZ modes on the flared cone. Mode 1 is the highest amplitude mode of the Markovian term $\Omega^{(0)}$, at $300khz$, which contains both the primary and secondary instability. Mode 2 is the highest amplitude mode of the Markovian term at the first higher harmonic ($600khz$). The memory modes $k=1,2,6$ are the highest amplitude modes of the memory term $\Omega^{(k)}$. These modes show a structure that is reminiscent of "hot" streaks that develop due to a so-called fundamental breakdown as observed in DNS. Notice the memory terms contain structures with larger amplitude present deeper into the turbulent region, and more complex structures within the transition region.}
\label{fig:mz_modes_on_cone}
\end{figure}

The unsteady wall pressure disturbance data obtained from DNS was used to extract the highest amplitude (and highest energy) Markovian modes (see \cref{fig:mz_modes_on_cone}). The location where the so-called "primary" streaks (see discussion in \cite{hader_fasel_2019} for details) appear and disappear in the time-averaged Stanton number contours on the surface of the cone  are marked with solid magenta lines in \cref{fig:mz_modes_on_cone}. Mode 1 in \cref{fig:mz_modes_on_cone} (top) corresponds to the highest amplitude term at $f=300$ kHz. This frequency is the dominant linear (primary) instability found in both DNS (\cite{hader_2018}) and experiments (\cite{chynoweth_2019}). The amplitude contours of mode 1 (\cref{fig:mz_modes_on_cone}) exhibit dominant axisymmetric structures in the upstream portion of the close up in \cref{fig:mz_modes_on_cone} (top) which suggests that these structures correspond to the dominant axisymmetric second mode waves. These axisymmetric structures begin to deform in the azimuthal direction near the location where the "primary" streaks begin to appear (see \cref{fig:mz_modes_on_cone}). The wavelength of this modulation corresponds to the spacing of the "hot" streaks observed in the Stanton number contours (\cref{fig:ch_contours}). Therefore, the Markovian mode also appears to capture the dominant secondary instability, which in the case of the flared cone, is a so-called fundamental breakdown where an axisymmetric large amplitude (primary) wave resonates with a lower amplitude oblique (secondary) wave with the same frequency. The amplitude distribution for $f=600$ kHz, first higher harmonic of the dominant primary instability, is plotted in \cref{fig:mz_modes_on_cone}. The amplitude contours clearly show that this higher harmonic is "activated" farther downstream compared to the primary wave with $300$ kHz. This is consistent with the understanding that this higher harmonic is nonlinearly generated by a self-interaction of the primary wave once sufficiently large amplitudes are reached. Initially the higher harmonic is also dominated by axisymmetric structures before an azimuthal modulation is again observed in the "primary" streak region (\cref{fig:mz_modes_on_cone}). The azimuthal wavelength corresponds again to the wavelength of the secondary wave undergoing the strongest resonance.

\begin{figure}[!htb]
\centering
\begin{subfigure}[b]{0.3\textwidth}
\centering
\includegraphics[width=1\textwidth]{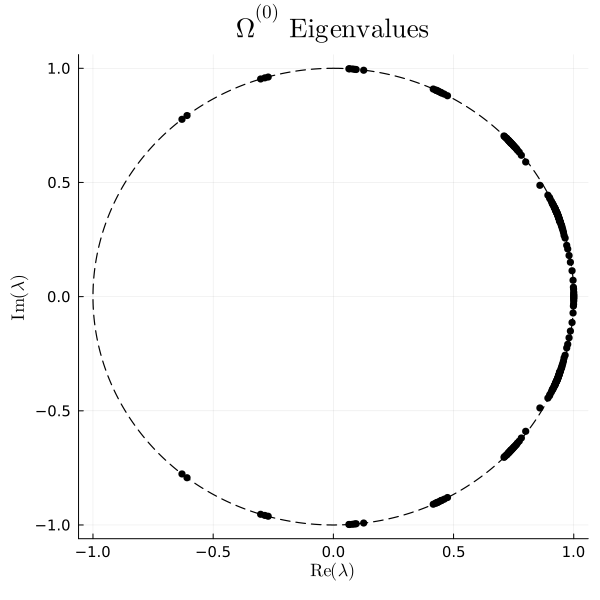}
\caption{Markovian term eigenvalues}
\end{subfigure}
\begin{subfigure}[b]{0.3\textwidth}
\centering
\includegraphics[width=1\textwidth]{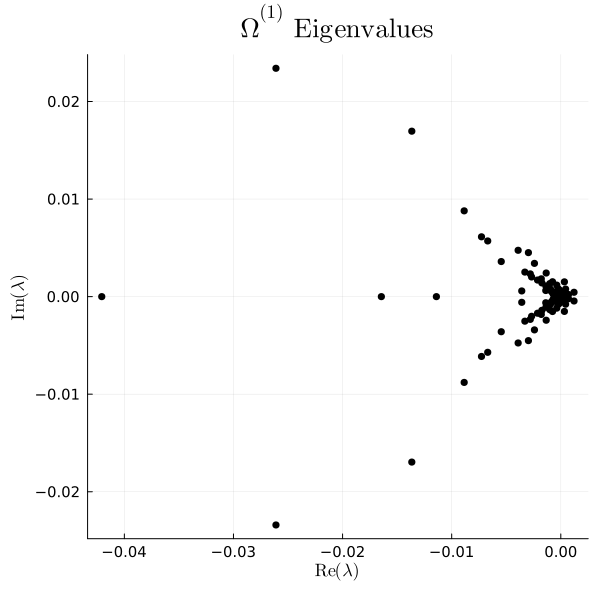}
\caption{$\Omega^{(1)}$ eigenvalues}
\end{subfigure}
\begin{subfigure}[b]{0.3\textwidth}
\centering
\includegraphics[width=1\textwidth]{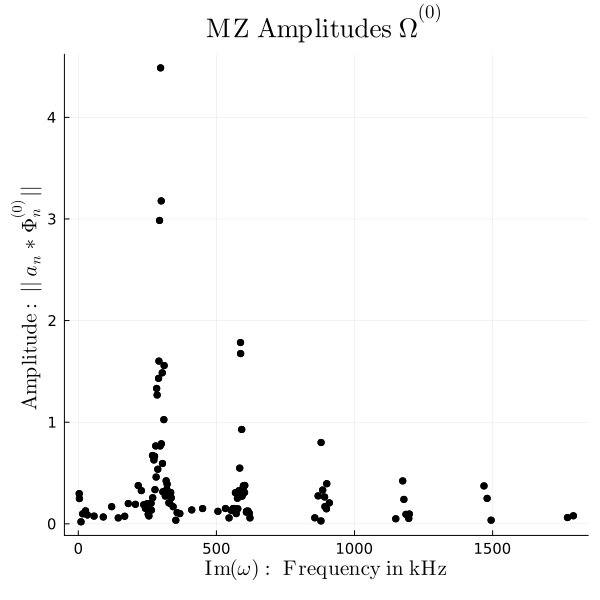}
\caption{Amplitude: modes of $\Omega^{(0)}$}
\end{subfigure}
\begin{subfigure}[b]{0.3\textwidth}
\centering
\includegraphics[width=1\textwidth]{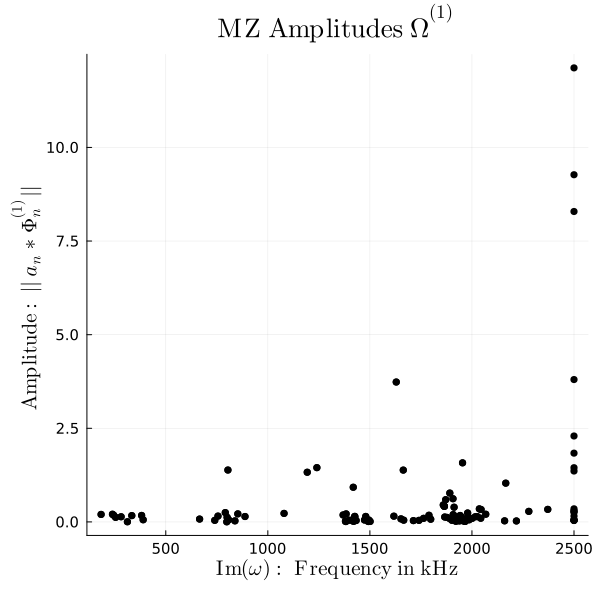}
\caption{Amplitude: modes of $\Omega^{(1)}$}
\end{subfigure}
\begin{subfigure}[b]{0.3\textwidth}
\centering
\includegraphics[width=1\textwidth]{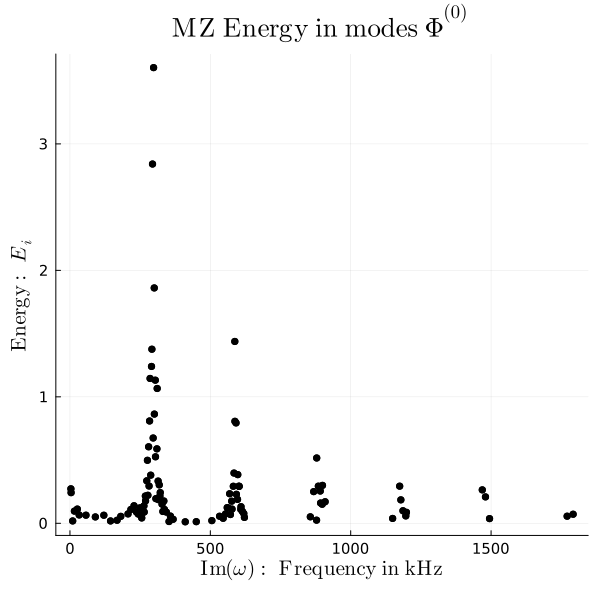}
\caption{Energy: modes of $\Omega^{(0)}$}
\end{subfigure}
\begin{subfigure}[b]{0.3\textwidth}
\centering
\includegraphics[width=1\textwidth]{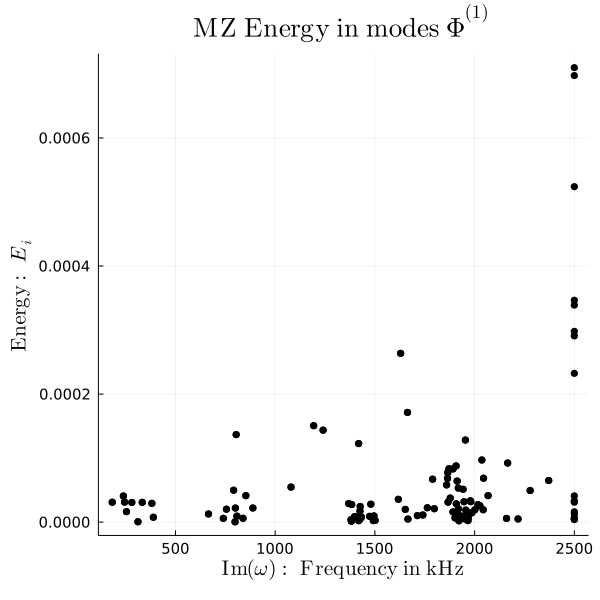}
\caption{Energy: modes of $\Omega^{(1)}$}
\end{subfigure}
\caption{(a, b) MZ Markovian and first memory term eigenvalues (c, d) Amplitudes of Markovian term and first memory term (e,f) energy contained in Markovian modes and memory modes.}
\label{fig:mz_modes_spectrum_energy}
\end{figure}

\begin{figure}[!htb]
\centering
\begin{subfigure}[b]{0.65\textwidth}
\centering
\includegraphics[width=1\textwidth]{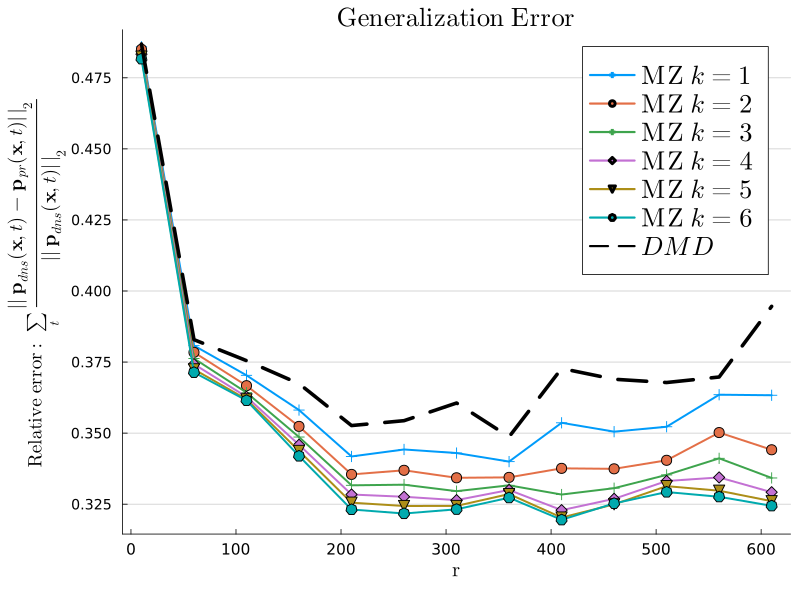}
\caption{Generalization error}
\end{subfigure}
\begin{subfigure}[b]{0.98\textwidth}
\centering
\includegraphics[width=1\textwidth]{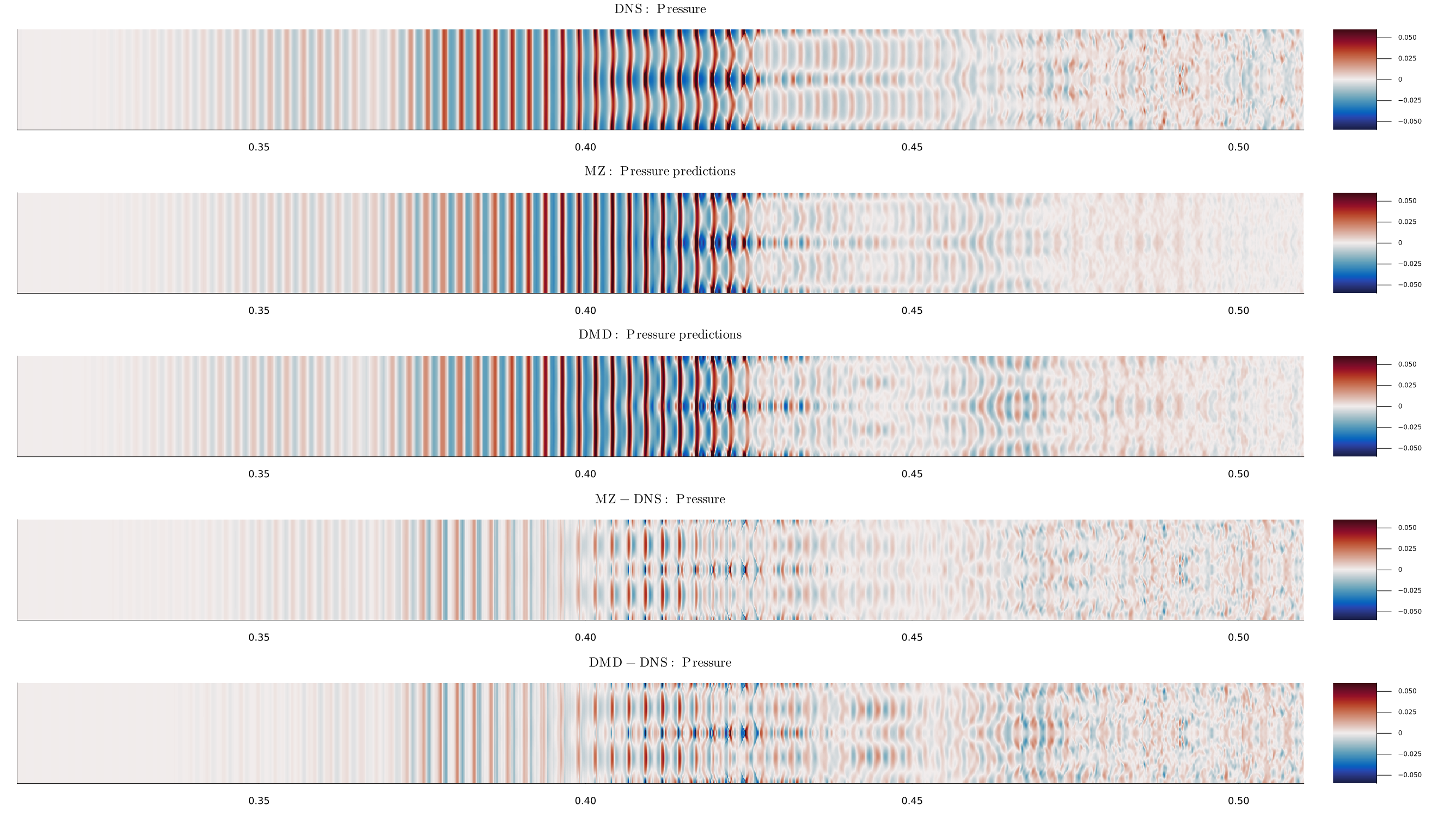}
\caption{Comparing pressure field of DNS (top) to MZ and DMD future state prediction (bottom) after 200 time steps}
\end{subfigure}

\caption{(a) Generalization error on future state predictions of the full pressure disturbance field extracted on the surface of the flared cone using Frobenius norm to compute the relative error over entire time frame. Comparing DMD vs MZ with different number of memory terms. As $r$ (the number of SVD modes used) increases, we see that the generalization error initially decreases, but then starts to increase.  In this case, the number of time steps used in forward state prediction is 250. We see that the relative error of MZ is a 5 to 15 percent improvement over DMD future state prediction. (b) Plotting a snapshot of MZ and DMD future state predictions and the difference of future state predictions with DNS of the pressure at the surface of the cone after 200 time steps (at the same region as plotted in Fig. \ref{fig:mz_modes_on_cone}). MZ and DMD are nearly able to predict the DNS pressure field even after 200 time steps (with $r=210$), although the DMD prediction has a larger error within the transition region, especially as seen in the  "hot" streaks (as seen in the error fields of the bottom two plots in (b)).}
\label{fig:prediction_comp}
\end{figure}

We see the MZ approach is able to extract nontrivial large scale spatio-temporal coherent structures of the Markovian term and memory effects which exhibits structures reminiscent of the "hot" streaks that develop due to the fundamental breakdown as observed in DNS \cite{hader_2018, hader_fasel_2019}. Furthermore, from Fig. \ref{fig:mz_modes_on_cone} we see that the large scale structures present in the memory terms, contain not only larger contributions from the turbulent region, but have nontrivial contributions to the transition region with clear imprints of the oblique waves, known to play an important role in the nonlinear generation of the "hot" streaks and ultimately transition to turbulence (see \cite{hader_fasel_2019}). Additionally, an eigenvalue analysis of MZ and DMD in Figs. \ref{fig:mz_modes_spectrum_energy} and \ref{fig:mz_modes_spectrum_energy_Atilde}, shows that these memory terms have nontrivial dynamical contributions as seen from the spectrum of each memory term. Finally, we make future state predictions with MZ and compare the relative error to DMD in Fig. \ref{fig:prediction_comp}, further demonstrating the dynamical relevance contained in the memory terms.

\begin{figure}[!htb]
\centering
\begin{subfigure}[b]{0.3\textwidth}
\centering
\includegraphics[width=1\textwidth]{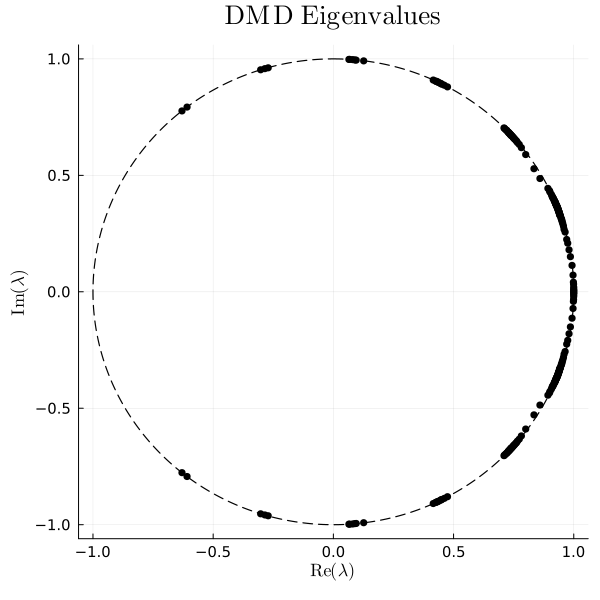}
\caption{Eigenvalues of DMD modes}
\end{subfigure}
\begin{subfigure}[b]{0.3\textwidth}
\centering
\includegraphics[width=1\textwidth]{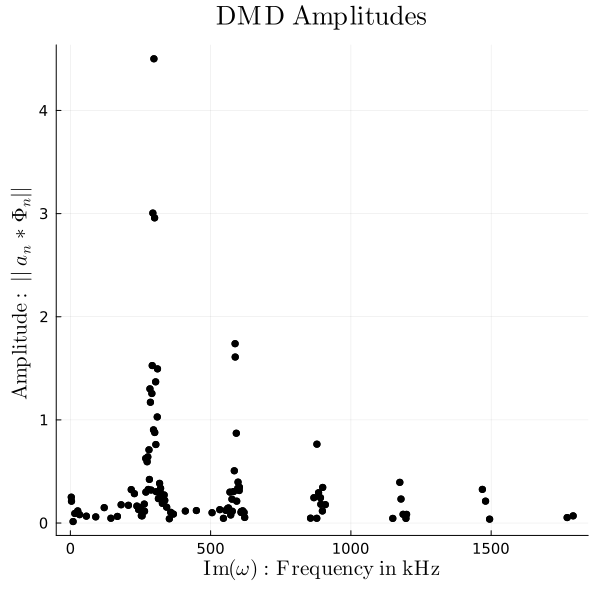}
\caption{Amplitude in DMD modes}
\end{subfigure}
\begin{subfigure}[b]{0.3\textwidth}
\centering
\includegraphics[width=1\textwidth]{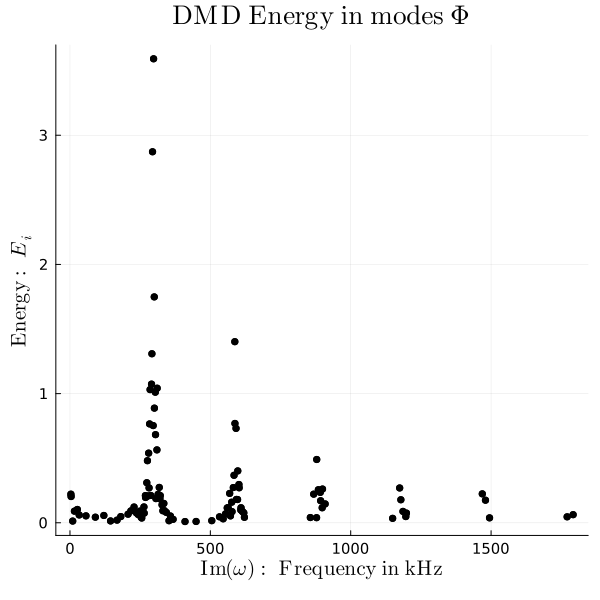}
\caption{Energy contained in DMD modes}
\end{subfigure}
\caption{(a) DMD eigenvalues (b) Amplitudes, and (c) Energy.}
\label{fig:dmd_spectrum_cone}
\end{figure}

\section{Conclusions and Future work}

In this manuscript, we have introduced a data-driven Mori-Zwanzig algorithm by using Mori's linear projector, and SVD based low rank approximation (equivalently projecting onto POD modes), and compared the results to DMD on both 2D flow over a cylinder and the pressure disturbance signal extracted on the surface of a flared cone from a high resolution 3D DNS of laminar-turbulent boundary-layer transition on a flared cone at Mach 6. With this framework, we showed that the modes and spectrum obtained from DMD are identical to the modes and spectrum of the Markovian operator of MZ. Additionally, by including more memory terms in the MZ framework, not only can this data-driven MZ formulation outperform DMD in future state prediction, but can also serve as a diagnostic tool to extract nontrivial large scale spatio-temporal structures of the memory effects. We showed that the memory terms play a significant role in the analysis of the laminar-turbulent
transition (Fig. \ref{fig:mz_modes_on_cone}), as well as increasing the accuracy of future state prediction over DMD by up to $15\%$ (Fig. \ref{fig:prediction_comp}). Thus, the data-driven MZ approach can serve as a way to understand both the Markovian term (similar to DMD) and non-Markovian structures of the flow. 

In the analysis of these coherent structures, it is the combination of both the Markovian and memory terms which illuminate identifiable structures present in the primary and secondary transition mechanisms relevant for this hypersonic flow. These structures, together with the spectrum and energy content contained in each mode, demonstrate the nontrivial Memory terms that contain structures present in the mechanisms generating the "hot streaks" seen in the flow. Furthermore, in addition to the Markovian modes that demonstrate the cascade from the low frequency from the primary instabilities to higher harmonics, the memory terms demonstrate an inverse cascade. This indicates that the memory terms improve the resolution in the transition region. We also showed that these memory terms contain nontrivial contributions to the primary and secondary mechanisms. 

There are many possible directions that future works can explore, such as investigating different selections of observables, nonlinear projections using regression \cite{lin22_nn_mz}, extracting 3D coherent structures from volumetric data, and analysis of the spectrum of the MZ operators in time-delay coordinates (Fig. \ref{fig:mz_modes_spectrum_energy_Atilde}). For example, how are the modes of the memory terms interacting with the modes of the Markovian term? This will be a direction of our future study.

\section{Acknowledgments}

This work has been co-authored by employees of Los Alamos National Laboratory (LANL), which
is operated by Triad National Security, LLC, for the National Nuclear Security
Administration of U.S. Department of Energy (Contract No. 89233218CNA000001). Funding was provided by the
LANL's LDRD program, project number 202200104DR.

\bibliographystyle{vancouver}
\bibliography{sci_tech_23.bib}

\appendix
\section{Algorithms}\label{sec:algs}

In Algorithm \ref{alg:obs} we first obtain the snapshot data as is done in POD and DMD but we also need to include some past history for which the parameter $k$ is used. 

\begin{algorithm}[H]
    \centering
    \caption{Discrete MZ Algorithm: SVD based observables}
    \begin{algorithmic}[1]
        \State Select the number of memory terms $k$ (typically $k \sim 10$)
        \State Given snapshots of data: $\bm X_{full} = [\bm x_1, ..., \bm x_{m+k}]$
        \State $\bm X_{full} \approx \bm U_r \bm \Sigma_r \bm V_r^*$ Truncated SVD
        \State Using SVD compressed observables (as linear autoencoder): i.e $\bm G = \bm U_r^* \bm X_{full}$ 
        
        \State Collect snapshots over $k$ time delays: $ \bm G_1 = [\bm g_1, \bm g_2, ..., \bm g_m]$, $ \bm G_2 = [\bm g_2, \bm g_2, ..., \bm g_{m+1}]$, ... $\bm G_{k} = [\bm g_{k}, \bm g_{k+1}, ..., \bm g_{m+k}]$ \\
        --------------------------------------------------------------------------------------------\\
        
        \vspace{0.5cm}
        $\bm C_1 = \left< \bm g(t), \bm g(t)^T \right> \approx \bm G_1 \cdot \bm G_1^T$
        \For{$i\gets 2, ...,k+1$}
        \State $\bm C_i = \left< e^{(i\Delta) \mathcal{L}} \bm g(t), \bm g(t)^T \right> \approx \bm G_{i} \cdot \bm G_{1}^T$
        \EndFor
        \State $\bm \Omega_1 = \bm C_2 \cdot \bm C_1^{-1}$
        \For{$i\gets 2, ...,k+1$}
        \State $\bm \Omega_i = \left[ \bm C_{i+1} - \sum_{l=1}^{i-1} \bm \Omega_l \cdot \bm C_{i-l+1} \right] \cdot \bm C_1^{-1}$
        \EndFor
        \begin{center}
            $\bm g_{n+1} = \sum_{l=1}^{k-1} \bm \Omega_l \bm g_{n-l} + \bm 0$ Future prediction in reduced space\\
            $\bm x_{n+1} = \bm U_r \bm g_{n+1}$ Future state prediction in full state space\\
            Extracting modes: $\bm \Phi^{(i)} = \bm U_r \bm W^{(i)}$, where $\bm \Omega_i \bm W^{(i)} = \bm W^{(i)} \bm \Lambda ^{(i)}$ is eigendecomposition. 
        \end{center}
    \end{algorithmic}
    \label{alg:obs}
\end{algorithm}

\begin{minipage}{0.48\textwidth}
\begin{algorithm}[H]
    \centering
    \caption{DMD Algorithm}\label{algorithm_dmd}
    \begin{algorithmic}[1]
        \State Given snapshots of full state observables: $ \bm X_1 = [\bm x_1, \bm x_2, ..., \bm x_m]$, $ \bm X_2 = [\bm x_2, \bm x_2, ..., \bm x_{m+1}].$ 
        \State Low rank approximation $\bm X_1 \approx \bm U_r \bm \Sigma_r \bm V_r^*$
        \State Project $\bm A$ onto POD modes:
        $\tilde{\bm A} = \bm U_r^* \bm A \bm U_r = \bm U_r^* \bm X_2 \bm V_r \bm \Sigma_r^{-1}$
        \State Compute eigendecomposition: $\tilde{\bm A} \bm W = \bm W \bm \Lambda$
        \State Eigenvectors of $\bm A$ given by $\bm \Phi = \bm X_2 \bm V_r \bm \Sigma_r^{-1} \bm W$ \hspace{1.5cm}
         ----------------------------------------------------------
        \begin{center}
            DMD reconstruction \\
            $\bm \Omega = log(\bm \Lambda) / dt$ \\
            $\bm b = \bm \Phi ^{\dagger} \bm g_1$ \\
            $\hat{\bm x}(t) = \bm b * \exp(\bm \Omega * t)$
        \end{center}

    \end{algorithmic}
\end{algorithm}
\end{minipage}
\hfill
\begin{minipage}{0.48\textwidth}
\begin{algorithm}[H]
    \centering
    \caption{Discrete MZ Algorithm: Low rank $\bm C$}\label{algorithm1}
    \begin{algorithmic}[1]
        \State Given snapshots of data: $\bm X_{full} = [\bm x_1, ..., \bm x_{m+k}]$
        \State $\bm X_{full} \approx \bm U_r \bm \Sigma_r \bm V_r^*$ Truncated SVD
        
        \State Collect snapshots over $k$ time delays: \\
        $ \bm X_1 = [\bm x_1, \bm x_2, ..., \bm x_m]$, $ \bm X_2 = [\bm x_2, \bm x_2, ..., \bm x_{m+1}]$, ... $\bm X_{k} = [\bm x_{k}, \bm x_{k+1}, ..., \bm x_{m+n\delta}]$ \\
        
        ------------------------------------------------------ \\
        
        \vspace{0.5cm}
        Project $\bm C$ onto POD modes $\bm C_1 \approx \bm U_r^* \bm X_1 \cdot \bm X_1^T \bm U_r$
        \For{$i\gets 2, ...,k+1$}
        \State $\bm C_i \approx \bm U_r^* \bm X_i \cdot \bm X_{1}^T \bm U_r$
        \EndFor
        \State $\bm \Omega_1 = \bm C_2 \cdot \bm C_1^{-1}$
        \For{$i\gets 2, ...,n\Delta+1$}
        \State $\bm \Omega_i = \left[ \bm C_{i+1} - \sum_{l=1}^{i-1} \bm \Omega_l \cdot \bm C_{i-l+1} \right] \cdot \bm C_1^{-1}$
        \EndFor
        \begin{center}
            MZ approximate reconstruction \\ 
            (orthogonal dynamics $\bm W_n = \bm 0$) \\
            $\bm g_{n+1} = \sum_{l=1}^{k-1} \bm \Omega_l \bm g_{n-l} + \bm 0$\\
            $\bm x_{n+1} = \bm U_r \bm g_{n+1}$\\
        \end{center}
    \end{algorithmic}
    \label{alg:low_rank}
\end{algorithm}
\end{minipage}

In Algorithm \ref{alg:low_rank} and \ref{alg:obs} are two different interpretations of the same procedure. Algorithm \ref{alg:obs} interprets the SVD as a linear auto-encoder as a automatic technique for selecting observables from data (as a lossy data compression to avoid the intractable computations of $\bm C_i = \bm X_i \cdot \bm X_1^T$) and algorithm \ref{alg:low_rank} interprets this as a low rank approximation of $\bm C_k$ by projecting onto the POD modes (which is what is done in the standard DMD algorithm) as can be seen in the side by side comparison. This equivalence can be seen by replacing $\bm G_k$ with $\bm U_r^* \bm X_k $ when computing $\bm C_k$.

\section{2D flow over a cylinder}\label{sec:2d_cyl}

\begin{figure}[!htb]
\centering
\begin{subfigure}[b]{0.3\textwidth}
\centering
\includegraphics[width=1\textwidth]{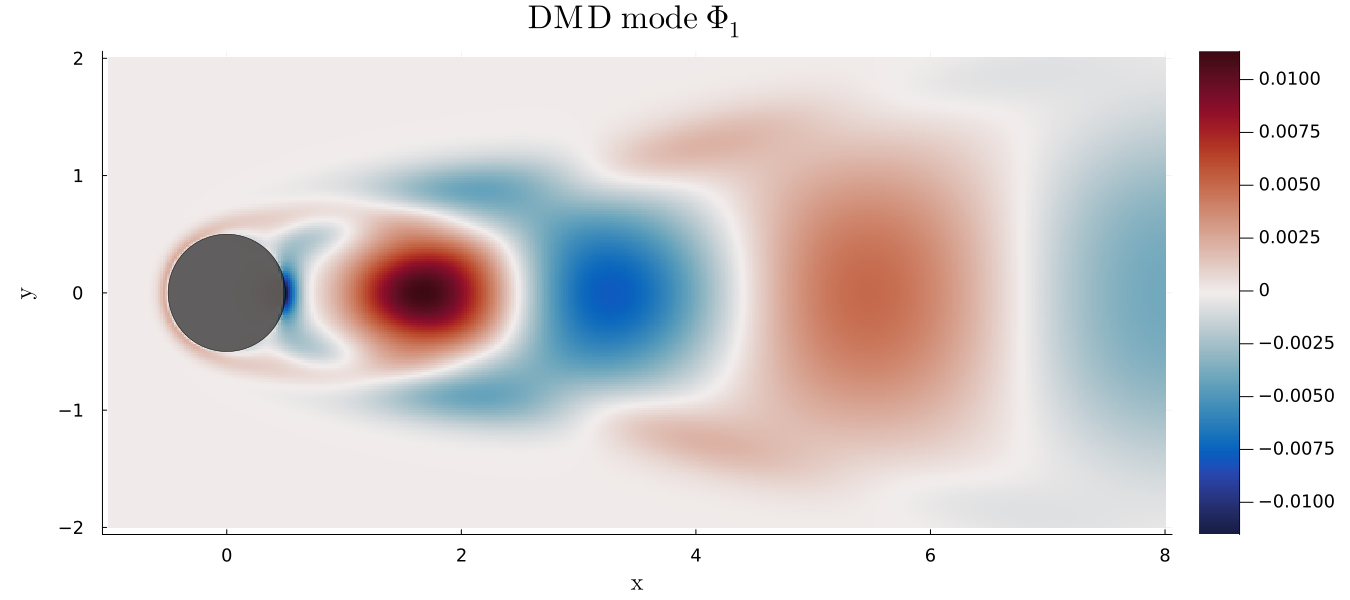}
\caption{DMD $\Phi_1$}
\end{subfigure}
\begin{subfigure}[b]{0.3\textwidth}
\centering
\includegraphics[width=1\textwidth]{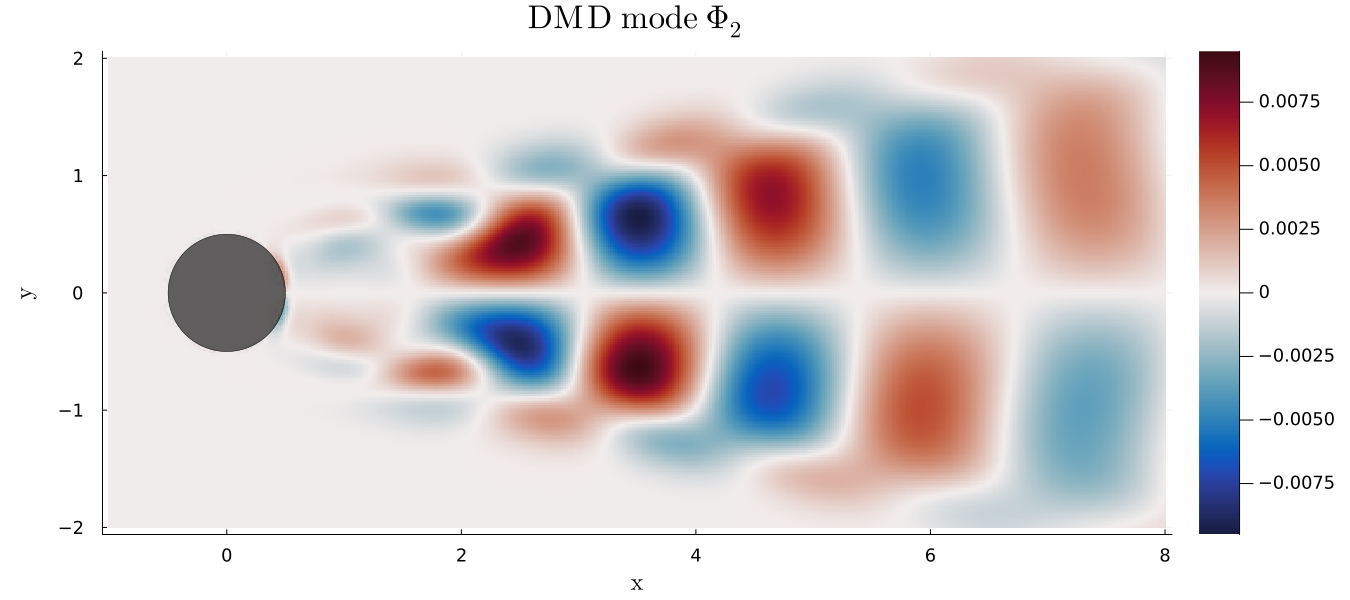}
\caption{DMD $\Phi_2$}
\end{subfigure}
\begin{subfigure}[b]{0.3\textwidth}
\centering
\includegraphics[width=1\textwidth]{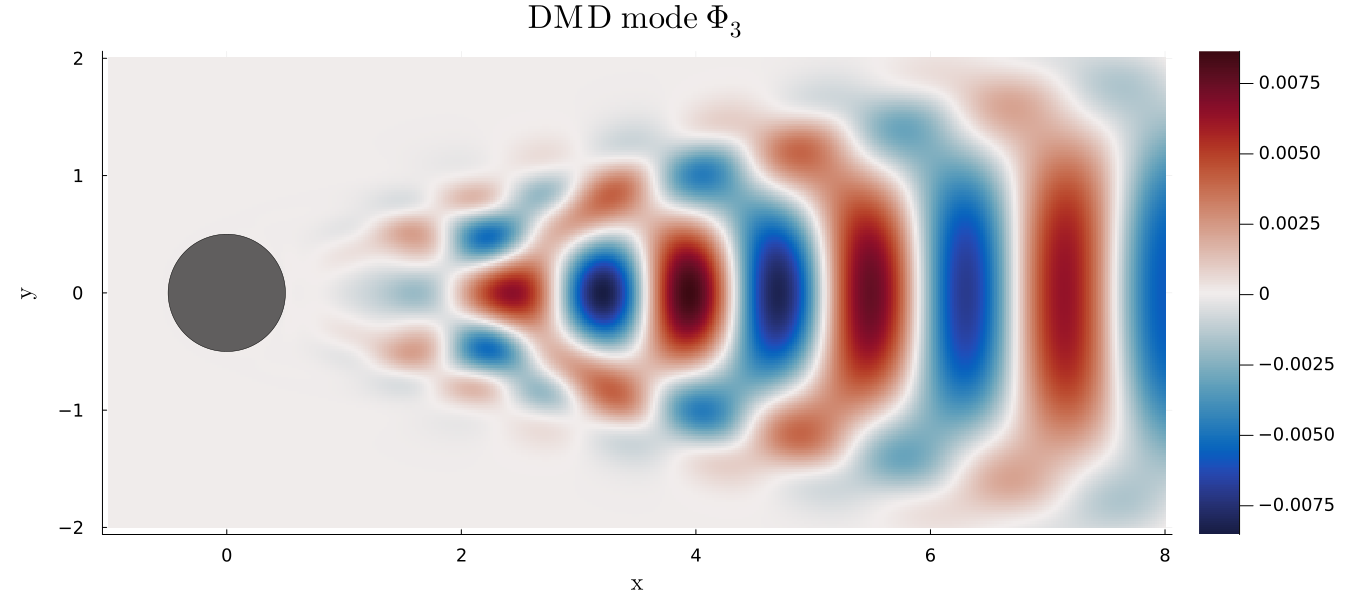}
\caption{DMD $\Phi_3$}
\end{subfigure}

\centering
\begin{subfigure}[b]{0.3\textwidth}
\centering
\includegraphics[width=1\textwidth]{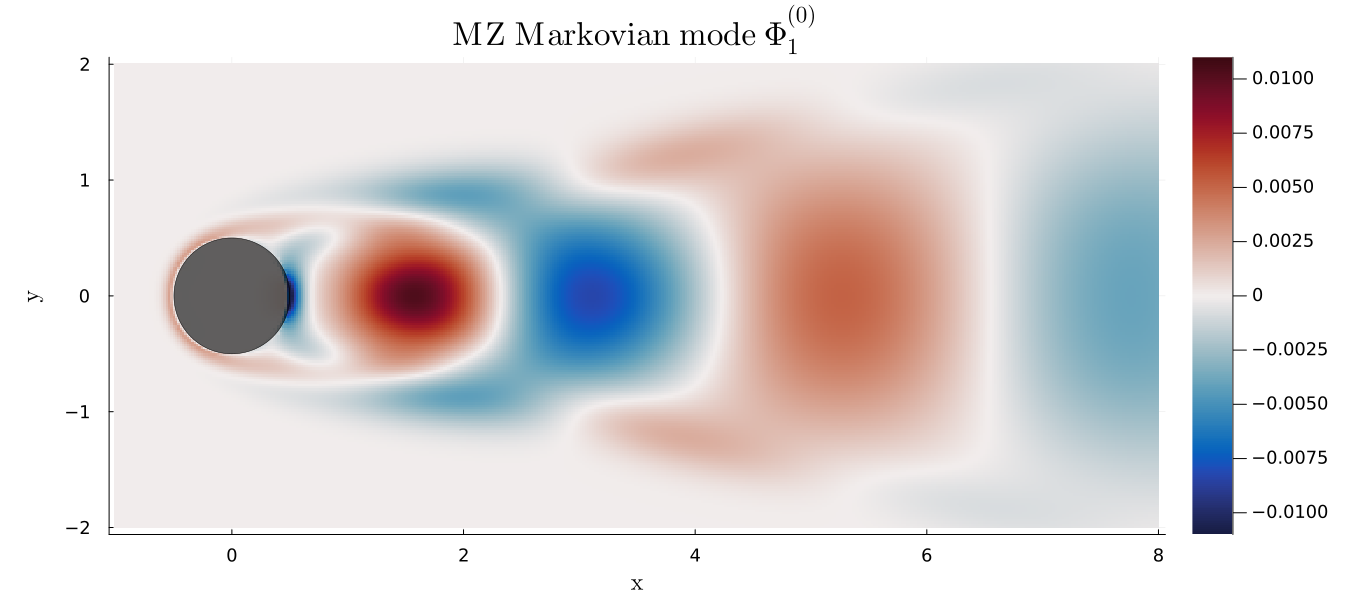}
\caption{MZ $\Phi^{(0)}_1$}
\end{subfigure}
\begin{subfigure}[b]{0.3\textwidth}
\centering
\includegraphics[width=1\textwidth]{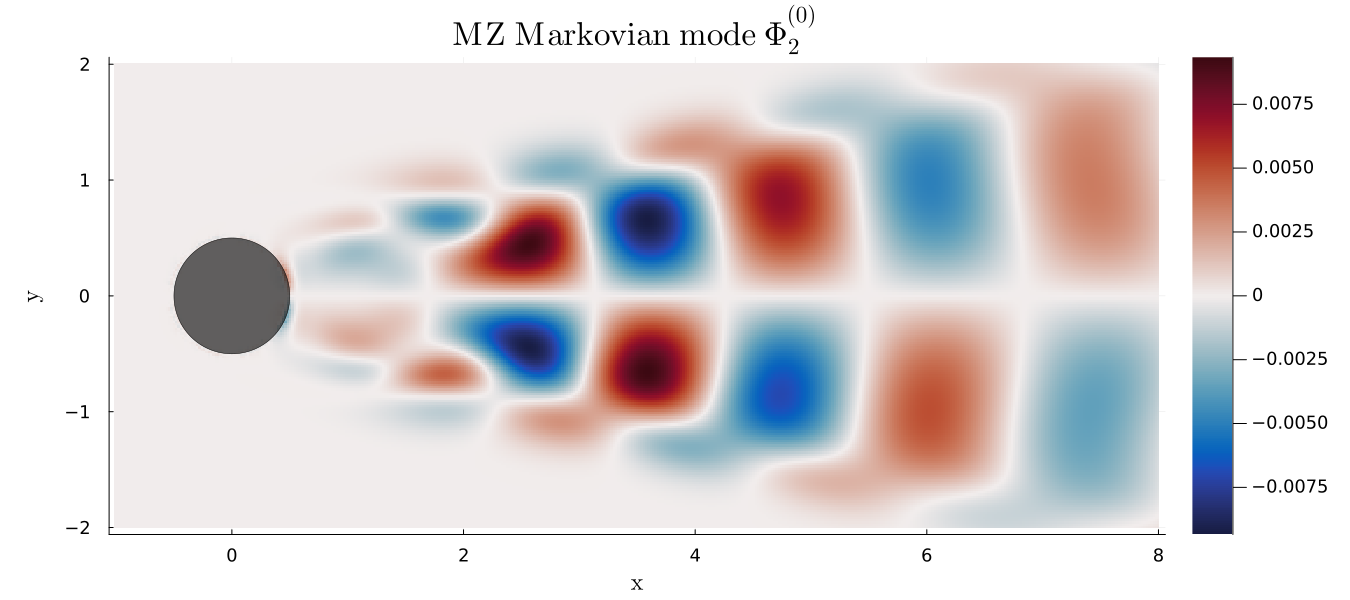}
\caption{MZ $\Phi^{(0)}_2$}
\end{subfigure}
\begin{subfigure}[b]{0.3\textwidth}
\centering
\includegraphics[width=1\textwidth]{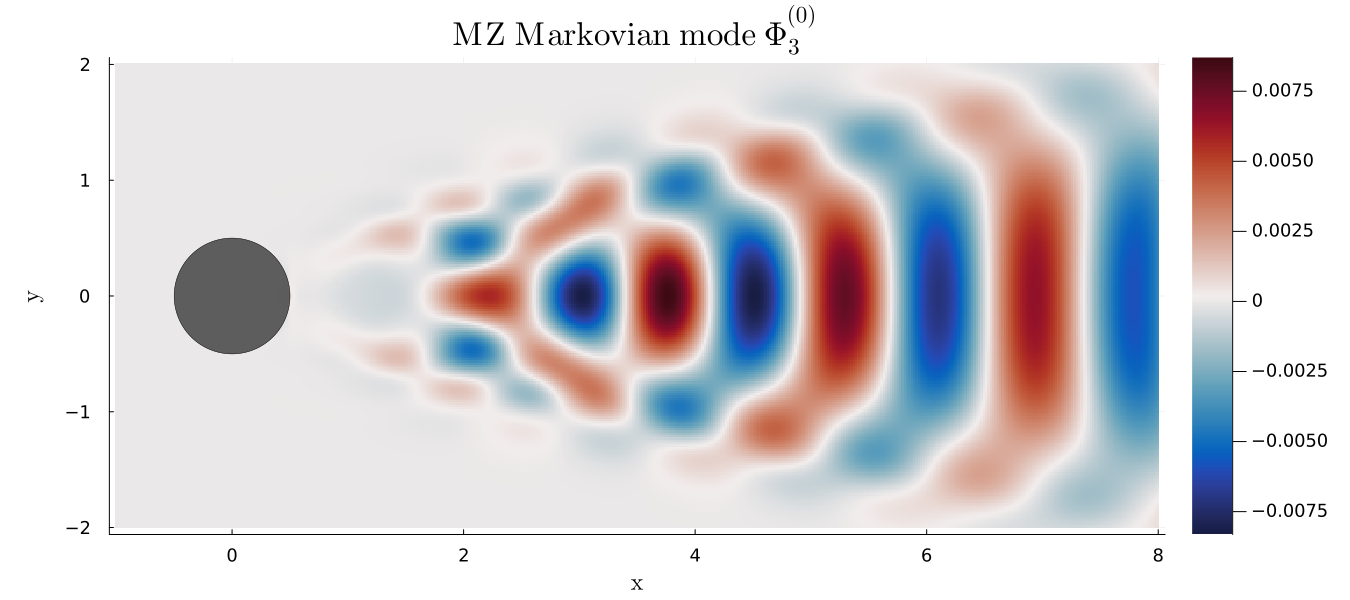}
\caption{MZ $\Phi^{(0)}_3$}
\end{subfigure}

\centering
\begin{subfigure}[b]{0.3\textwidth}
\centering
\includegraphics[width=1\textwidth]{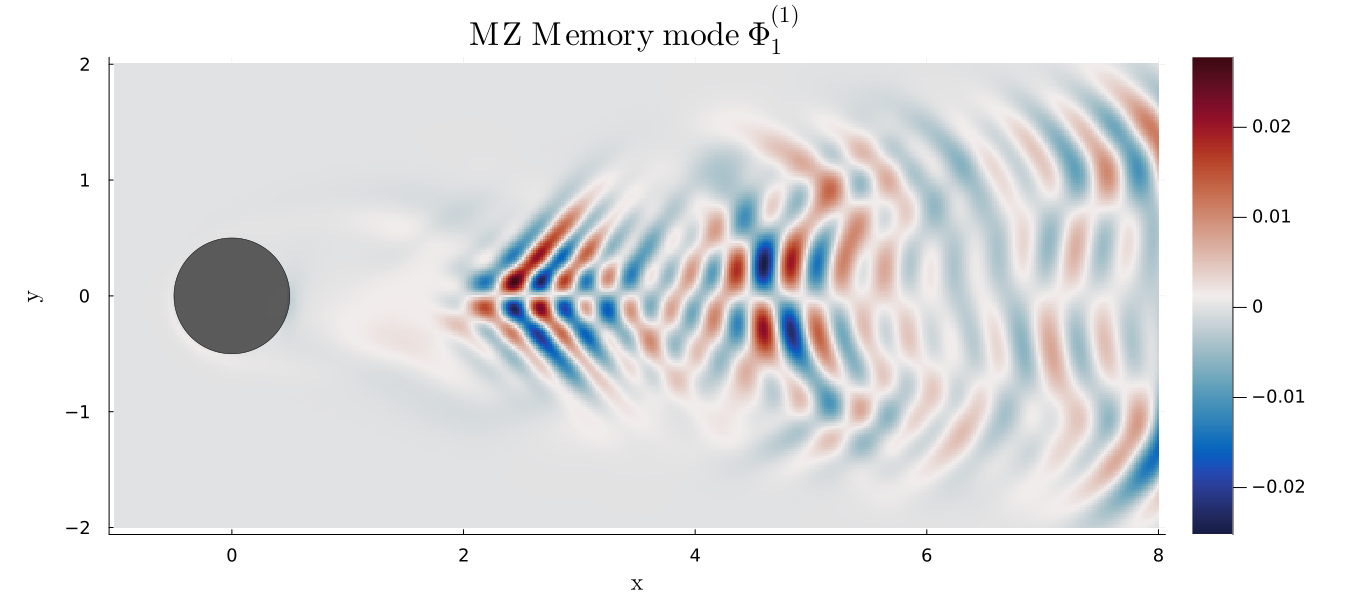}
\caption{MZ $\Phi^{(1)}_1$}
\end{subfigure}
\begin{subfigure}[b]{0.3\textwidth}
\centering
\includegraphics[width=1\textwidth]{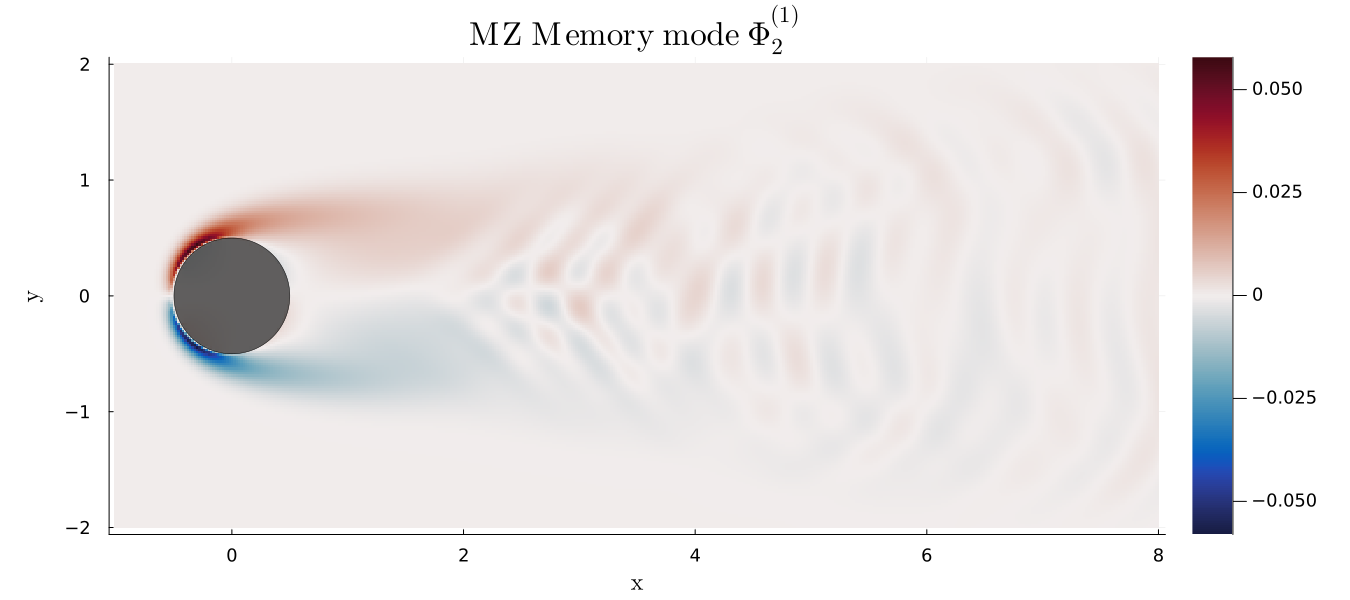}
\caption{MZ $\Phi^{(1)}_2$}
\end{subfigure}
\begin{subfigure}[b]{0.3\textwidth}
\centering
\includegraphics[width=1\textwidth]{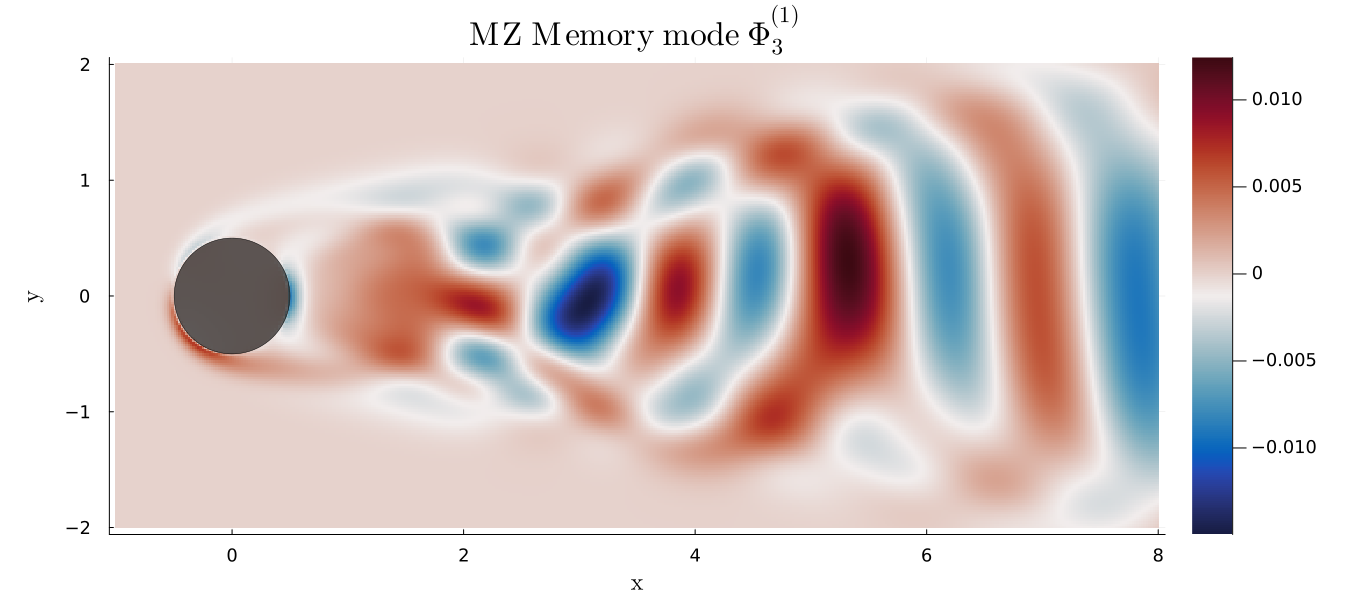}
\caption{MZ $\Phi^{(1)}_3$}
\end{subfigure}

\centering
\begin{subfigure}[b]{0.3\textwidth}
\centering
\includegraphics[width=1\textwidth]{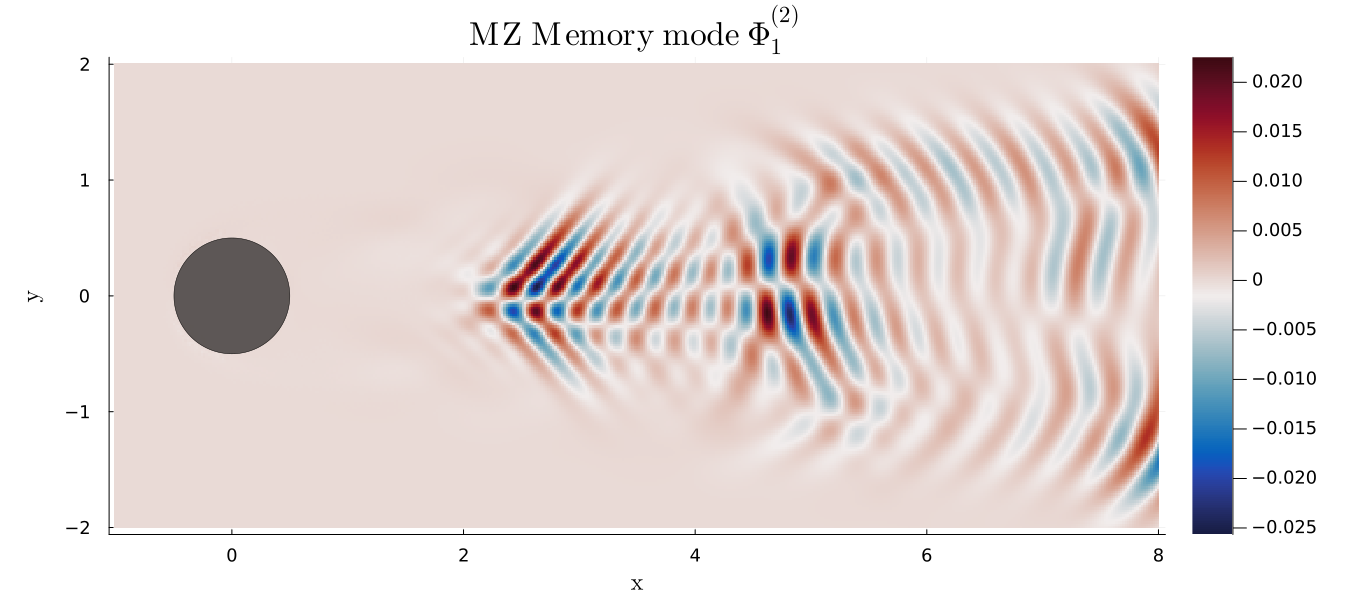}
\caption{MZ $\Phi^{(2)}_1$}
\end{subfigure}
\begin{subfigure}[b]{0.3\textwidth}
\centering
\includegraphics[width=1\textwidth]{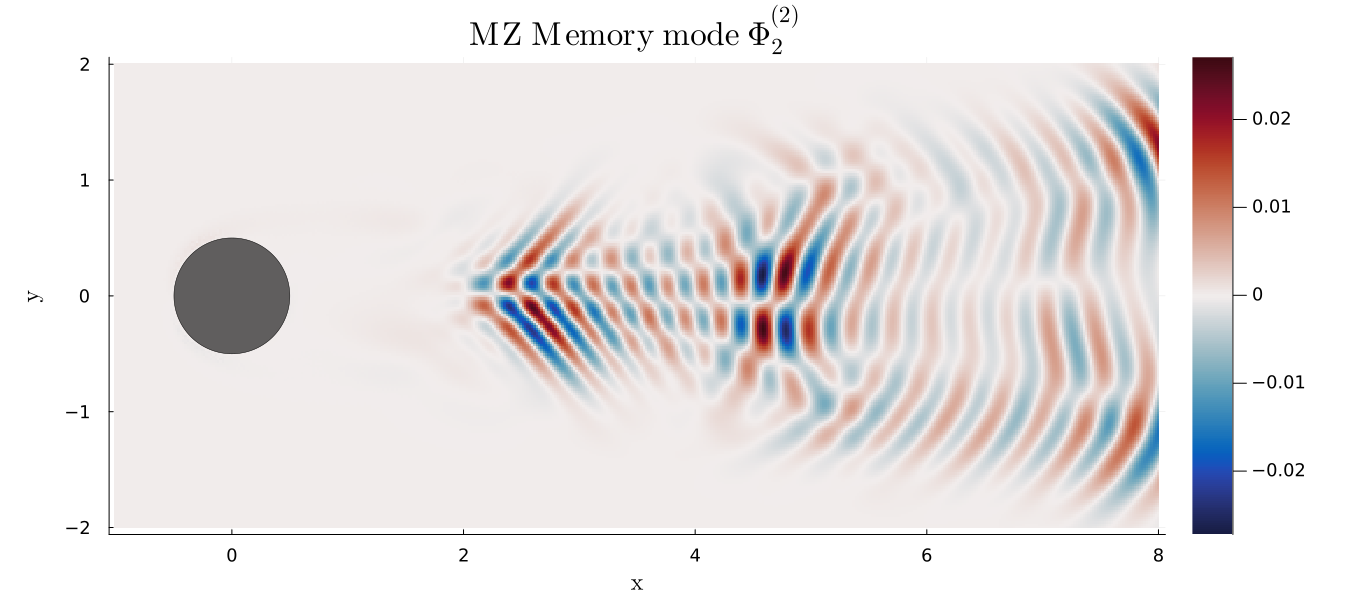}
\caption{MZ $\Phi^{(2)}_2$}
\end{subfigure}
\begin{subfigure}[b]{0.3\textwidth}
\centering
\includegraphics[width=1\textwidth]{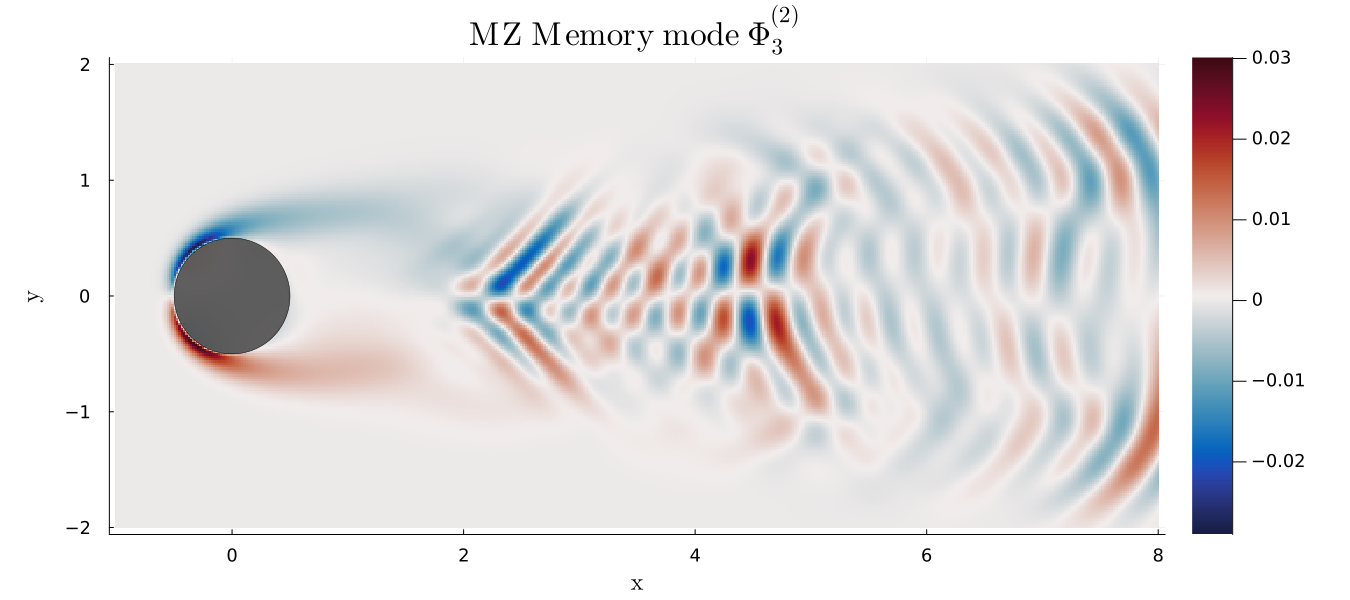}
\caption{MZ $\Phi^{(2)}_3$}
\end{subfigure}
\caption{MZ modes of Markovian, 1st and 2nd memory terms vs DMD modes for 2D flow over a cylinder ad the structure of the first memory (where vorticity data was used). We see that the modes of the Markovian term and DMD modes are identical and the 1st memory term has non-trivial structure (although nearly trivial temporal relevance as seen below).}
\label{fig:dmd_vs_mz_modes}
\end{figure}

\begin{figure}[!htb]
\centering
\begin{subfigure}[b]{0.42\textwidth}
\centering
\includegraphics[width=1\textwidth]{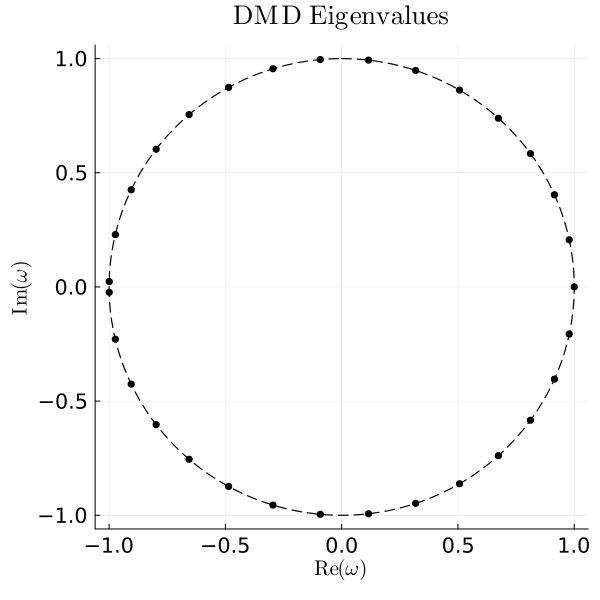}
\caption{DMD: Eigenvalues}
\end{subfigure}
\begin{subfigure}[b]{0.42\textwidth}
\centering
\includegraphics[width=1\textwidth]{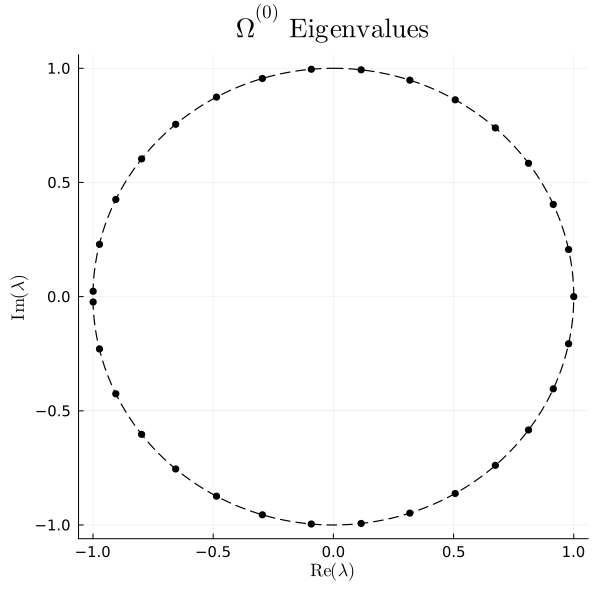}
\caption{MZ Markovian term: Eigenvalues}
\end{subfigure}
\begin{subfigure}[b]{0.42\textwidth}
\centering
\includegraphics[width=1\textwidth]{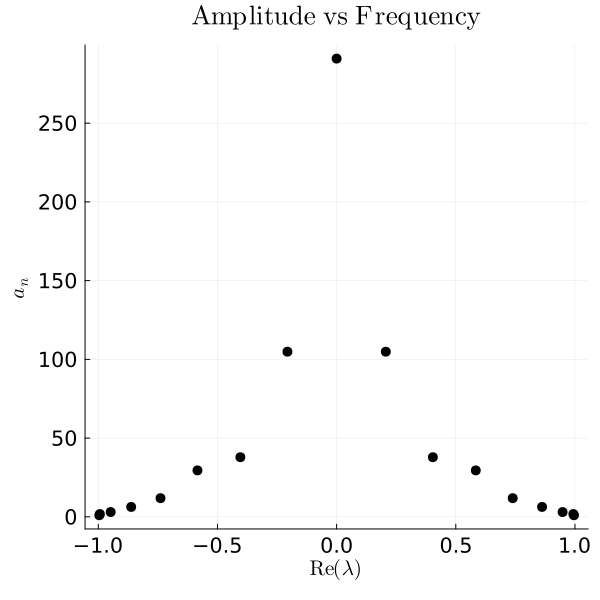}
\caption{DMD: Spectrum}
\end{subfigure}
\begin{subfigure}[b]{0.42\textwidth}
\centering
\includegraphics[width=1\textwidth]{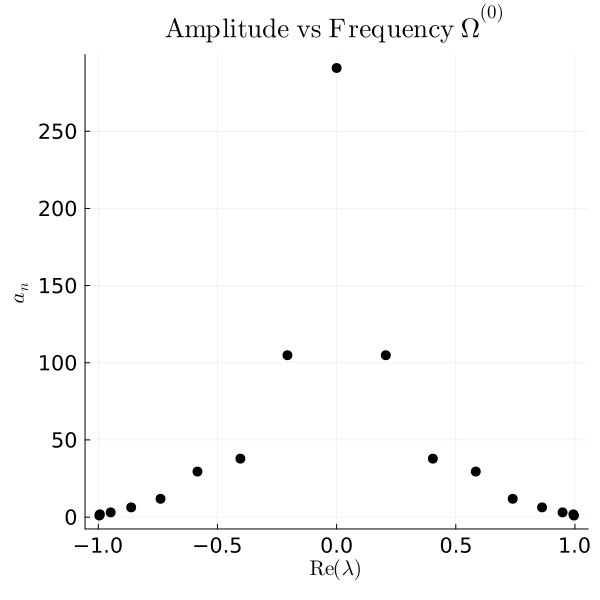}
\caption{MZ Markovian term: Spectrum}
\end{subfigure}
\centering
\begin{subfigure}[b]{0.36\textwidth}
\centering
\includegraphics[width=0.95\textwidth]{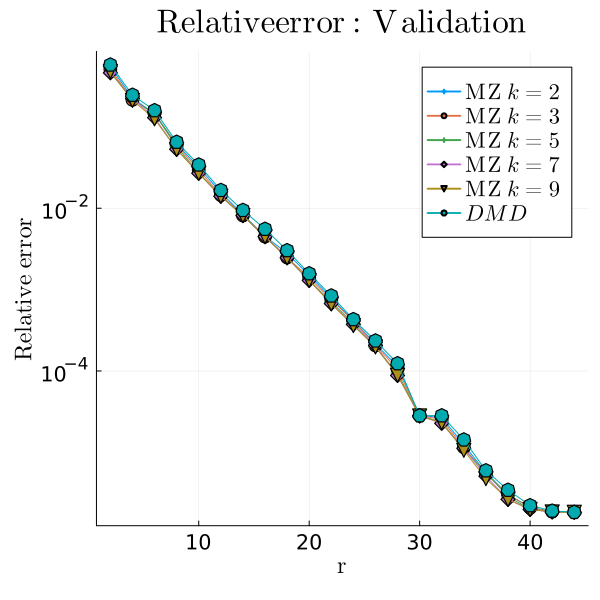}
\caption{MZ and DMD generalization errors}
\end{subfigure}
\caption{(a,b,c,d) Comparing Eigenvalues and amplitudes of modes from DMD and Markovian term shows similar temporal dynamics. (e) Shows the generalization errors computed by measuring relative errors of future state predictions of MZ and DMD models show that for this simple flow adding memory terms has a small but non-trivial improvement (f) on future state predictions. }
\label{fig:dmd_vs_mz_spectrum}
\end{figure}

2D flow over a cylinder is a standard data set used to test modal analysis tools. In this case we use $Re = 100$, resulting in vortex shedding (see \cite{dmd_book} and references therein for data sets).  In this context, it is used as a simple problem to compare DMD vs MZ on their ability to serve as a modal analysis tool as well as a ROM. We demonstrate in \autoref{fig:dmd_vs_mz_modes} that the modes of the Markovian term in MZ are nearly identical to that of the DMD modes. However, additional information is gained through the non-trivial structure identified in the memory kernels which quantifies how the unresolved variables, such the truncated modes and the other flow variables $u,v$, are effecting the resolved variables. However, in this low Reynolds number flow, it is also observed that, the Markovian term is enough to represent the flow, as seen in the memory terms, which although contain nontrivial spatial structures have relatively low dynamical relevance to the flow as demonstrated with the spectrum and generalization error \autoref{fig:dmd_vs_mz_spectrum}. Furthermore, in \autoref{fig:dmd_vs_mz_spectrum} we show the temporal dynamics of these modes are similar by comparing eigenvalues of $\bm \Omega^{(0)}$ and $\bm A$ as well as the spectrum.

\FloatBarrier

\subsection{Analysis of MZ operators for hypersonic flow}

In this section we investigate and analyze the contributions from the MZ operators to the dynamics through the companion matrix $\tilde{\mathbf{A}}$.  Using time-delay coordinates, the discrete time GLE can be written of the form $\tilde{\mathbf{g}}_{n+1} = \tilde{\mathbf{A}} \tilde{\mathbf{g}}_{n} $. Where, 

$$ \tilde{\mathbf{A}} = \begin{bmatrix}
\mathbf{\Omega^{(0)}} & \mathbf{\Omega^{(1)}}  & ... & \mathbf{\Omega^{(k)}}\\
\mathbf{I_r} & \mathbf{0} & ... & \mathbf{0}\\
\vdots &  \ddots &  &  \vdots \\
\mathbf{0} & ... &  \mathbf{I_r} & \mathbf{0}\\
\end{bmatrix}$$

and the time-delay coordinates are $$ \tilde{\mathbf{g}}_{n} = \begin{bmatrix}
\mathbf{g}_{n} \\
\mathbf{g}_{n-1} \\
\vdots \\
\mathbf{g}_{n-k} \\
\end{bmatrix} .$$

\begin{figure}[!htb]
\centering

\begin{subfigure}[b]{0.3\textwidth}
\centering
\includegraphics[width=1\textwidth]{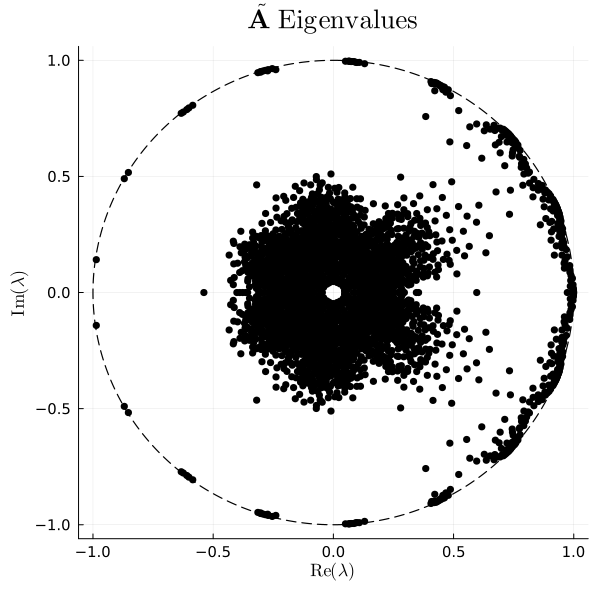}
\caption{$\bm \tilde{A}$ eigenvalues}
\end{subfigure}
\begin{subfigure}[b]{0.3\textwidth}
\centering
\includegraphics[width=1\textwidth]{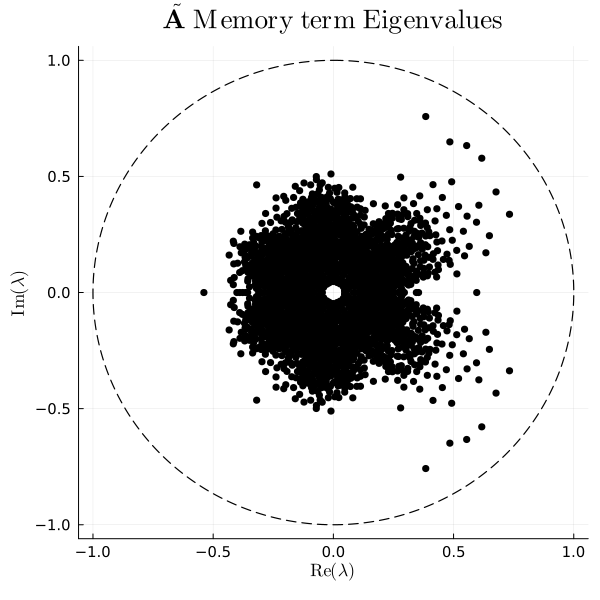}
\caption{$\bm \tilde{A}$ eigenvalues: memory}
\end{subfigure}
\begin{subfigure}[b]{0.3\textwidth}
\centering
\includegraphics[width=1\textwidth]{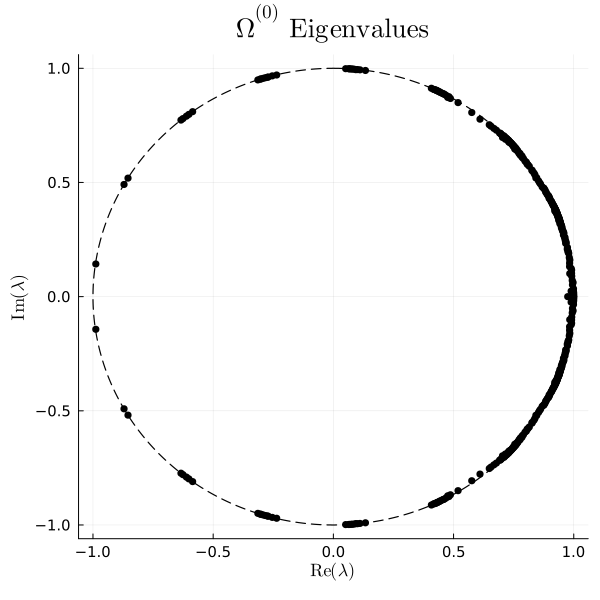}
\caption{Markovian term alone}
\end{subfigure}
\caption{(a, b) $\bm \tilde{A}$ eigenvalues which contains the contribution from the Markovian and memory terms. (c) Eigenvalues of Markovian term alone.}
\label{fig:mz_modes_spectrum_energy_Atilde}
\end{figure}

Fig. \ref{fig:mz_modes_spectrum_energy_Atilde} shows there is an interaction between the spectrum of the individual MZ operators in the long time dynamics characterized by the companion form. This interaction is an investigation for future works.

\end{document}